\documentclass[letterpaper]{article} 
\usepackage{aaai25}  
\usepackage{times}  
\usepackage{helvet}  
\usepackage{courier}  
\usepackage[hyphens]{url}  
\usepackage{graphicx} 
\usepackage{natbib}  
\usepackage{caption} 
\usepackage{algorithm}
\usepackage{algorithmic}
\usepackage{enumerate}
\usepackage{enumitem}
\usepackage{tabularx}
\usepackage{caption}
\usepackage{todonotes}
\usepackage{subcaption}
\usepackage{booktabs}
\usepackage{soul}
\usepackage{multirow}
\usepackage{makecell}
\usepackage{array}
\usepackage{amsmath}
\usepackage{xspace}
\usepackage{xcolor}
\usepackage{float}
\usepackage{placeins}
\usepackage{tikz}
\newcommand{\answerYes}[1]{\textcolor{blue}{#1}} 
 
\newcommand{\answerNA}[1]{\textcolor{gray}{#1}}

\usepackage{newfloat}
\usepackage{listings}
\DeclareCaptionStyle{ruled}{labelfont=normalfont,labelsep=colon,strut=off} 
\floatstyle{ruled}
\newfloat{listing}{tb}{lst}{}
\floatname{listing}{Listing}
\title{The Susceptibility Paradox in Online Social Influence}
\author{
    Luca Luceri\equalcontrib,
    Jinyi Ye\equalcontrib,
    Julie Jiang,
    Emilio Ferrara
}
\affiliations{
    Information Sciences Institute, University of Southern California\\
    Thomas Lord Department of Computer Science, University of Southern California\\
    lluceri@isi.edu, jinyiy@usc.edu, yioujian@usc.edu, emiliofe@usc.edu
}

\usepackage{bibentry}

\begin{document}

\maketitle

\begin{abstract}
    Understanding susceptibility to online influence is crucial for mitigating the spread of misinformation and protecting vulnerable audiences. This paper investigates susceptibility to influence within social networks, focusing on the differential effects of influence-driven versus spontaneous behaviors on user content adoption. Our analysis reveals that influence-driven adoption exhibits high homophily, indicating that individuals prone to influence often connect with similarly susceptible peers, thereby reinforcing peer influence dynamics, whereas spontaneous adoption shows significant but lower homophily. Additionally, we extend the \emph{Generalized Friendship Paradox} to influence-driven behaviors, demonstrating that users' friends are generally more susceptible to influence than the users themselves, \textit{de facto} establishing the notion of \textit{Susceptibility Paradox} in online social influence. This pattern does not hold for spontaneous behaviors, where friends exhibit fewer spontaneous adoptions. We find that susceptibility to influence can be predicted using friends' susceptibility alone, while predicting spontaneous adoption requires additional features, such as user metadata. These findings highlight the complex interplay between user engagement and characteristics in spontaneous content adoption. Our results provide new insights into social influence mechanisms and offer implications for designing more effective moderation strategies to protect vulnerable audiences.
\end{abstract}

%

\section{Introduction}
On social media, social influence plays a significant role in disseminating information and shaping beliefs, attitudes, and behaviors. Politically, platforms like Twitter and Facebook are arenas for spreading biased information and fake news, which can potentially disrupt the integrity of voting events \cite{gonzalez2023asymmetric,stella2018bots}. Economically, brands can sway consumers' purchasing decisions through targeted advertising and influencer partnerships \cite{aral2012identifying,bakshy2012social}. Socio-culturally, social media is pivotal in accelerating social movements and activism \cite{den2007ideologically,mundt2018scaling}.


Research has shown that online social influence can be a powerful tool for persuading behavioral change, especially when messages are tailored to specific audiences \cite{oyibo2019relationship}. Identifying groups highly susceptible to such influences is, therefore, essential. While many studies have examined various factors that contribute to susceptibility to influence—such as personality traits \cite{oyibo2019relationship}, political leanings \cite{calvillo2021individual}, information credibility \cite{traberg2022birds}, and exposure frequency \cite{hassan2021effects}—they often overlook the relational aspect and networked nature of influence dynamics. Consequently, the interconnectedness and potential peer influence within networks are not fully accounted for understanding and predicting susceptibility. For instance, how an individual's likelihood of being influenced relates to the susceptibility levels of their social connections remains underexplored.

\subsection{Contributions of This Work}

In this paper, we present an empirical framework designed to understand, model, and predict users' susceptibility to influence on social media platforms. This framework takes into account the networked nature of social influence, with a focus on homophilous influence dynamics, and investigates how the \emph{Generalized Friendship Paradox} manifests in influence-driven versus spontaneous behaviors, thereby establishing the concept of \emph{Susceptibility Paradox}.
In detail, we address the following Research Questions (RQs):
\begin{enumerate}[label={RQ\arabic*}:,align=left]
\item \textit{Is susceptibility to influence homophilous within a social network?} We aim to investigate whether a user's susceptibility is correlated with the susceptibility of their friends.
\item\textit{Does the \emph{Generalized Friendship Paradox} extend to user susceptibility to influence?} Our objective is to explore whether a user's friends are typically more susceptible to influence than the user themselves.
\item \textit{How effectively can a user's susceptibility be predicted by their friends' susceptibility?} We seek to measure to what extent a user's level of susceptibility can be predicted using information from their social network.
\end{enumerate}

By leveraging two large-scale Twitter datasets covering political and public health discussions, our study reveals that susceptibility to influence is homophilous within social networks. We demonstrate that influence-driven behaviors exhibit greater homophily than spontaneous behaviors. Additionally, we find that users' friends are generally more susceptible to influence and less prone to spontaneous actions than the users themselves, supporting that the \emph{Generalized Friendship Paradox} holds for influence-driven behavior but not for spontaneous action in friendship networks. These insights inform predictive models, which can predict users' rate of influence-driven adoptions by leveraging only their friends' susceptibility to influence. In contrast, spontaneous behaviors are more difficult to predict using solely friends' influence signals, suggesting a complex interplay of user engagement and preferences in spontaneous actions.

Our findings are consistent across the two large-scale Twitter datasets and provide crucial insights into diverse behaviors, vulnerabilities, and influence dynamics on social media platforms. This knowledge can inform network interventions designed to protect vulnerable audiences and mitigate risks associated with influence operations, propaganda, and misinformation campaigns.

\section{Related Work}

\subsection{Susceptibility to Social Influence}
Numerous studies have investigated the factors that make individuals susceptible to influence, categorizing these factors as either personal or external. From a \textit{personal} standpoint, demographics such as age and gender play significant roles, with younger individuals and men often being more susceptible to influence when adopting new products \cite{aral2012identifying}. Personality traits like neuroticism, openness, and conscientiousness are also recognized as consistent predictors of susceptibility to influence strategies \cite{oyibo2019relationship}. In the political sphere, individuals with right-leaning ideologies are more susceptible to misinformation and deception \cite{calvillo2021individual,luceri2019red,pennycook2019lazy}, particularly those who exhibit lower trust in scientific authority or a tendency towards conspiracy thinking \cite{saling2021no}.

\textit{External} factors also play a role in susceptibility. Individuals are more likely to accept claims from sources they perceive as credible \cite{traberg2022birds}, and this tendency is further amplified when the persuasive message aligns with their partisanship or ideological views \cite{guess2018selective,moravec2018fake}. Additionally, the ``illusory truth" effect occurs when repeated exposure to specific information makes a claim appear more believable, regardless of its veracity, thereby increasing its acceptance \cite{hassan2021effects}.

\paragraph{Influence and Network Dynamics. }

A user network may be an under-explored external factor that affects an individual's susceptibility to social influence. Individuals with similar characteristics often cluster within networks, leading to homophilous patterns of content exposure and adoption behaviors \cite{mcpherson2001birds, aral2012identifying}. Furthermore, the structure of these networks often enhances the visibility of information among similar users, potentially fostering echo chambers \cite{cinelli2021echo}. These effects are intensified by algorithmic recommendations that tailor content exposure to match the prevailing interests within one's network \cite{guess2020exposure}. Such biased exposure can contribute to the ``majority illusion,'' where individuals mistakenly believe that the views or behaviors prevalent in their immediate social circle are representative of the general population \cite{lerman2016majority}. A structural cause of this phenomenon is known as the ``friendship paradox,'' which we will explore in the following section.

\subsection{The \emph{Generalized Friendship Paradox}}
Network structure can skew local observations. The friendship paradox states that, on average, your friends have more friends than you do \cite{feld1991your}. This occurs because high-degree nodes (individuals with many friends) are more likely to be sampled when considering the friends of a random individual. The friendship paradox is more pronounced in networks with heavy-tailed degree distributions, where a small number of nodes have a large number of connections \cite{lerman2016majority}.

The \emph{Generalized Friendship Paradox} (GFP) extends the original concept to node attributes beyond the number of friends. It posits that the characteristics of the neighbors of a node in any network will, on average, be greater than those of the node itself \cite{eom2014generalized}. For example, \citet{hodas2013friendship} examined the GFP in the context of online social networks, showing that users' friends are generally more active and receive more viral content on platforms like Twitter. Additionally, \citet{higham2019centrality} found that individuals' friends are more central in social networks.


Research also suggests that the GFP could bias local perceptions and individuals' understanding of societal norms \cite{alipourfard2020friendship}, potentially amplifying or diminishing the spread of information and behaviors. \citet{christakis2013social} highlight that the adoption of health-related habits, such as smoking, drinking, and exercising, is influenced by perceptions of peers' actions. Similarly, an individual's likelihood of adopting information may be influenced by exposure to friends' sharing behaviors (such as the frequency and type of content they share), and might increase when content is popular within an ego network. 
Although this foundational work provides valuable insights into influence dynamics on social media, the relationship between a user's susceptibility to influence and that of their friends remains underexplored. In this work, our aim is to fill this gap by investigating how network structural properties---specifically homophily and the GFP---shape patterns of susceptibility to influence within social networks.



\section{Methodology}

\subsection{Framework for Susceptibility Modeling}
Susceptibility to online influence is commonly modeled in probabilistic terms, indicating the likelihood that a user's behavior and/or beliefs may be influenced by online social interactions \cite{zhou2019network,romero2011differences,goyal2010learning,luceri2018deep}. The existing literature consistently highlights two primary components of susceptibility to influence: \textit{Exposure} to content and the potential subsequent \textit{Adoption} of that content \cite{aral2012identifying, hoang2016tracking}. We adhere to this formulation to construct our \textit{Susceptibility Framework}, which aims to estimate users' susceptibility to influence based on the dynamics between exposure and adoption.

\paragraph{Exposure.} 
On social media, exposure to a piece of content typically refers to a user encountering or viewing the content. Measuring exposure from observational social media data can be challenging, as detailed user behavioral data, such as scrolling and clicking, remain largely inaccessible to academics. Additionally, it is increasingly difficult to get access to the structure of the user follower network and, therefore, to whose content each user is exposed to from those whom they follow.
The situation is further complicated by \textit{algorithmic feed curation}, i.e., recommendation algorithms displaying potentially relevant content to a user that does not originate from the users they are following \cite{guess2020exposure}. Despite these challenges, several studies have approximated exposure by analyzing user interactions \cite{ferrara2015measuring, rao2022partisan, sasahara2021social, ye2023susceptibility}. The core assumption is that any form of engagement—whether retweeting, quoting, or replying—indicates a user's potential exposure to another user's content. These interactions imply that the user has not only viewed but also interacted with the content, thereby increasing the likelihood of subsequent exposures to that user's content. The rationale is to leverage past interactions as a proxy to predict future exposure.

Therefore, we operationalize exposure as follows: If a user $u_{t}$ interacts with another user $u_{s}$ through a retweet, quote, or reply, we consider $u_{t}$ exposed to all tweets posted by $u_{s}$ \textit{after} their first observed interaction.

\begin{figure}
    \centering    
    \includegraphics[width=\columnwidth]{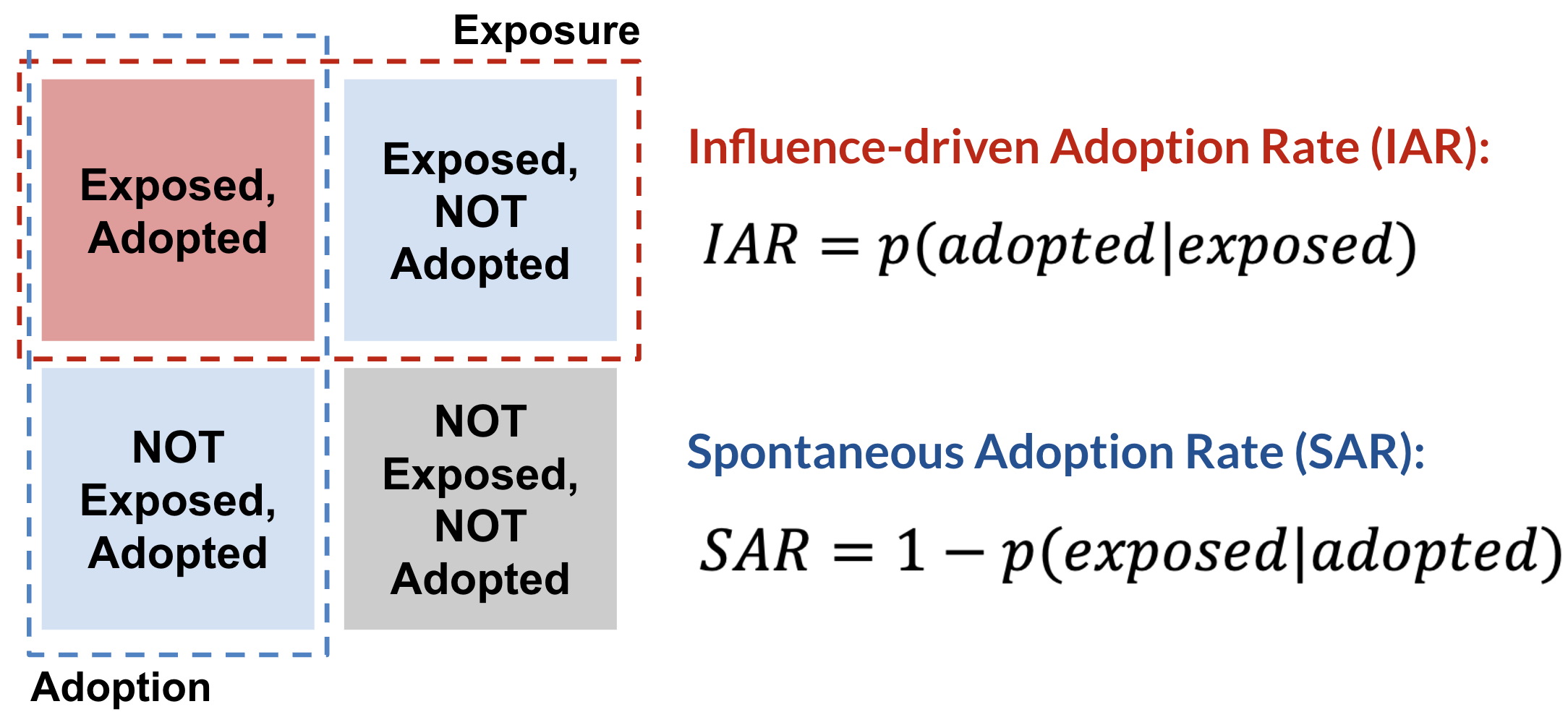} 
    \caption{An illustration of our \textit{Susceptibility Framework}. }
    \label{metrics}
\end{figure}

\paragraph{Adoption.} 
In this study, we define adoption as the action of sharing a URL, whether through an original tweet, a retweet, a quoted tweet, or a reply. We focus on URLs because they are commonly embedded in tweets and correspond to specific pieces of information, including multimodal content, such as images and videos, allowing for straightforward tracking of their diffusion and adoption within a social network.
Sharing a URL, whether through original content or interactions, represents a form of information adoption, reflecting a user's active decision to disseminate specific content. However, it is crucial to note that sharing or adopting a URL does not necessarily imply trust or endorsement of its content. Although retweeting is often seen as an endorsement \cite{boyd2010tweet, metaxas2015retweets}, other sharing activities can serve to question or critique content \cite{hemsley2018tweeting}.

\paragraph{Susceptibility Metrics.}
Building on our conceptualization of \textit{Exposure} and \textit{Adoption}, we develop a framework to estimate user \textit{susceptibility} to influence (see Figure \ref{metrics}). Specifically, by analyzing user activity, we evaluated two crucial dimensions: potential exposure to content and potential adoption of content. It is important to acknowledge that our data is ultimately an incomplete rendition of the real world: There may be instances where users have shared or been exposed to content not included in our dataset.  

Henceforth, we classify the relationship between a user and a piece of content into four distinct scenarios:
\begin{enumerate}[label=(\roman*),align=left]
    \item \textit{Exposed and adopted}: the user viewed and then shared the content;
    \item \textit{Exposed but not adopted}: the user viewed the content but did not share it;
    \item \textit{Not exposed but adopted}: the user shared the content without any prior exposure;
    \item \textit{Neither exposed nor adopted}: the user neither viewed nor ever shared the content.
\end{enumerate}


Based on this framework, and in line with the definitions provided by \cite{aral2012identifying}, we categorize adoption into two main types:  \textit{influence-driven adoption}, where the content is shared post-exposure, and \textit{spontaneous adoption}, where the content is shared without any prior exposure. Recognizing these distinctions is crucial, particularly since adoptions can occur independently of direct influence from others.
Motivated by this categorization, we introduce two susceptibility metrics: the \textit{Influence-Driven Adoption Rate} (IAR) and the \textit{Spontaneous Adoption Rate} (SAR). 

\textit{\textbf{Influence-driven Adoption Rate (IAR).}} This metric calculates the ratio of URLs that a user adopted after being exposed to them to the total number of URLs the user was exposed to, yielding a value between $0$ and $1$. Mathematically, it is expressed as:
\begin{align}
\textbf{IAR} = p(\text{adopted} \mid \text{exposed}) \nonumber = \frac{|E_{u_t} \cap A_{u_t}|}{|E_{u_t}|},
\end{align}
where $E_{u_t}$ denotes the set of all tweets containing URLs to which the user $u_t$ was exposed, and $A_{u_t}$ denotes the set of all tweets containing URLs that the user $u_t$ actually adopted. Following the methodology in \cite{bakshy2012role,ye2023susceptibility}, we implement a buffer period of two months for $A_{u_t}$. This period counts adoptions only if they occur after the first two months of our dataset, ensuring there is enough time to observe potential exposures before any subsequent adoptions.

\textit{\textbf{Spontaneous Adoption Rate (SAR).}} This metric measures the proportion of URLs that a user adopts without prior exposure, with values ranging from $0$ to $1$. It is mathematically expressed as:
\begin{align}
\textbf{SAR} = 1- p(\text{exposed} \mid \text{adopted}) \nonumber = 1 - \frac{|E_{u_t} \cap A_{u_t}|}{|A_{u_t}|},
\end{align}
where $E_{u_t}$ and $A_{u_t}$ are defined similarly to the IAR metric.

The distributions of the IAR and SAR metrics for both datasets are shown in the \textit{Appendix} (Figure \ref{IAR-SAR-frequency}). The strong negative Spearman correlation between IAR and SAR (\textsc{Election} dataset: $\rho = -0.855$, $\textit{p} < 0.001$; \textsc{Covid} dataset: $\rho = -0.857$, $\textit{p} < 0.001$) indicates that users who frequently adopt content after being exposed to it (i.e., high IAR) tend to adopt content without prior exposure less frequently (i.e., low SAR), and vice versa. This observation supports the validity of our susceptibility framework and particularly these two metrics, in capturing distinct behaviors on social media—whether users are more influenced by external sources (high IAR) or more prone to independent (i.e., spontaneous) content sharing (high SAR). It is important to note that a high susceptibility level is typically associated with a high IAR and a low SAR. We maintain these two distinct metrics to more accurately reflect the nuances of influence dynamics and user behaviors.

\subsection{Data Collection and Curation}

\begin{table}
  \centering
  \small
  \begin{tabular}{lcccc}
  \toprule
    & \multicolumn{2}{c}{\textsc{Election}} & \multicolumn{2}{c}{\textsc{Covid}}\\
    \midrule
    All tweets w/ URL & \multicolumn{2}{r}{64M} & \multicolumn{2}{r}{76M}\\
    Total users & \multicolumn{2}{r}{4.5M} & \multicolumn{2}{r}{10.8M}\\
    Target users & \multicolumn{2}{r}{0.7M} & \multicolumn{2}{r}{1.0M}\\
    \midrule
    Original tweets & \multicolumn{2}{r}{7.2M} & \multicolumn{2}{r}{18.8M}\\
    Retweets & \multicolumn{2}{r}{27.5M} & \multicolumn{2}{r}{38.4M}\\
    Quoted tweets & \multicolumn{2}{r}{18.5M} & \multicolumn{2}{r}{10.9M}\\
    Reply tweets & \multicolumn{2}{r}{10.8M} & \multicolumn{2}{r}{7.9M}\\
    \midrule
    $G(V,E)$ & $|V|$ & $|E|$ & $|V|$ & $|E|$\\
    \cmidrule(lr){2-3} \cmidrule(lr){4-5}
    Interaction & 49,794 & 117,384 & 73,960 & 163,974\\
    Retweet & 25,077 & 49,679 & 53,161 & 90,800\\
    Mention & 45,289 & 72,554 & 56,060 & 81,969\\
    \bottomrule
\end{tabular}
  \caption{Number of users and tweets, along with the number of nodes ($|V|$) and edges ($|E|$) in the friendship network $G(V,E)$.}
  \label{dataset}
\end{table}

To ensure the robustness of our findings across different contexts, we conduct our experiments on two datasets from Twitter (now \textit{X}), referred to as the \textsc{Election} dataset and the \textsc{Covid} dataset. These datasets focus on two major events: the 2020 US Presidential Election and the COVID-19 global pandemic, covering a comprehensive range of topics related to political and public health discussions. 

The \textsc{Election} dataset \cite{chen2021election2020} was collected beginning January 1st, 2020, and the \textsc{Covid} dataset \cite{chen2020tracking} beginning January 21st, 2020. For this research, we utilize data from the 12-month period of the same year, ending on December 31, 2020. Both datasets were collected via Twitter's streaming API, which gathered a  
sample of all tweets using a list of manually curated, topic-specific keywords, accounts, and hashtags. The tweets were fetched in real-time, ensuring data completeness and avoiding well-known re-hydration issues \cite{pfeffer2023just}. 

Since our methodology relies on tracking users' exposed to and adoption of information via URLs embedded in shared tweets, in this study, we only include tweets that contain at least one URL. Detailed statistics on the number of tweets with URLs, the number of engaged users, and the distribution of tweet categories 
are provided in Table \ref{dataset}.

\paragraph{Selecting ``Target Users".}
We further refine our dataset to isolate a specific subset of \textit{target users}. In line with the methodology of \citet{nikolov2021right}, we establish a minimum inclusion criterion: we study only users who shared at least 10 URLs of any kind during the overall observation period (12 months). This threshold ensures that we focus on users with sufficient data to reliably assess their susceptibility. Applying this criterion, our data narrows down to 698,808 users for the \textsc{Election} dataset and 1,047,566 users for the \textsc{Covid} dataset. 

\begin{figure}
    \centering    
    \includegraphics[width=8cm]{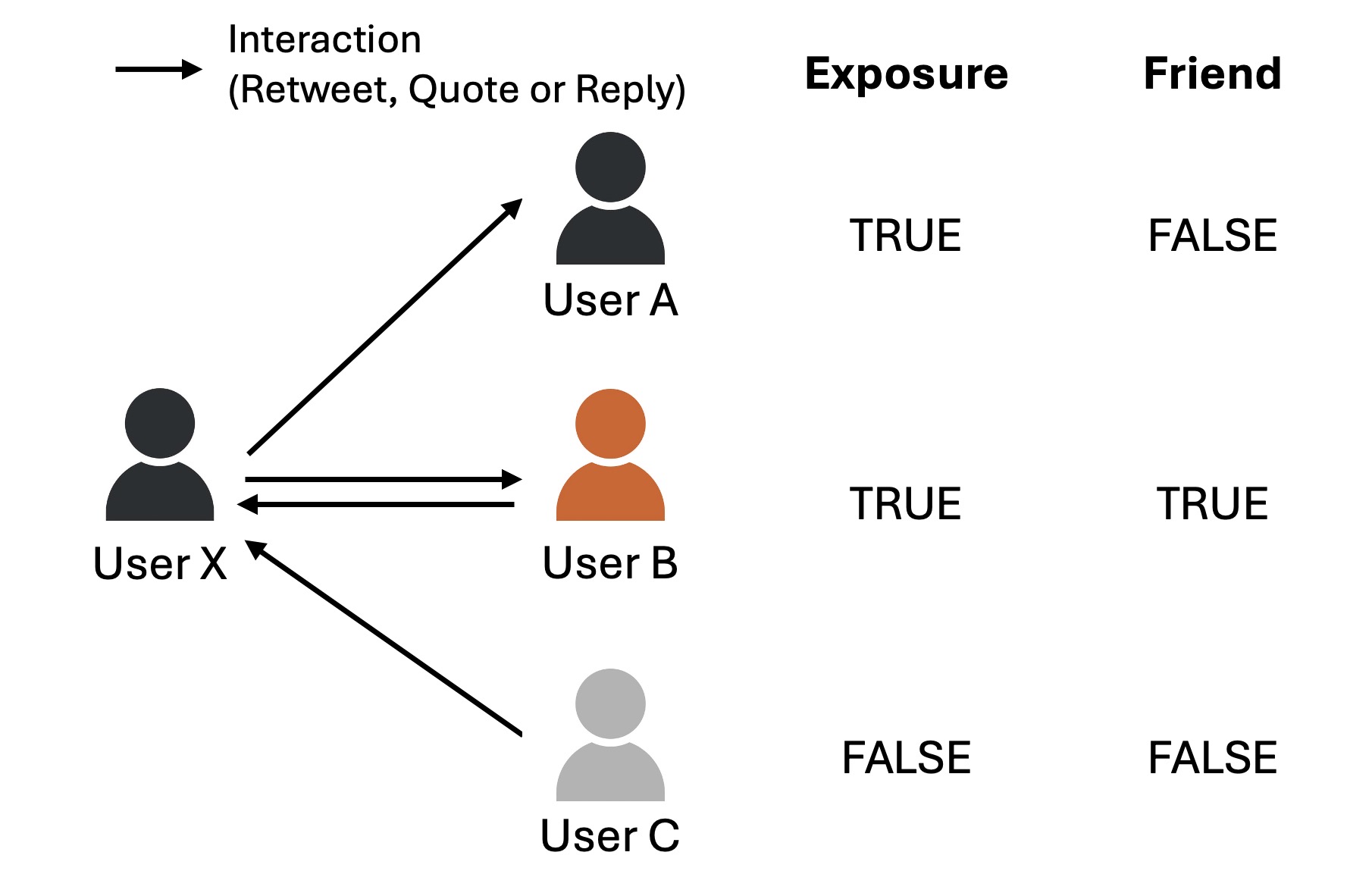} 
    \caption{Differentiation between exposure and friendship. For User X, the unidirectional interaction with User A is modeled as \textit{exposure}, while the bidirectional interaction with User B is modeled as \textit{friendship}.}
    \label{exposure_friend}
\end{figure}

\paragraph{Identifying ``Friends'' and Constructing Friendship Networks.}
Friendship networks are social structures formed by interconnected individuals who are linked through mutual recognition, shared activities, interactions, or sentiments \cite{feld1981focused,
zeggelink1996emergence}. We define two target users as ``friends'' if they engage in active reciprocal interactions. This means both users need to actively engage with the other via retweets, quotes, or replies. 
Although the \textit{followers} network could provide a more robust definition of friendship, such data is challenging to obtain using the Twitter API \cite{martha2013study}. In lieu of the \textit{followers} data, we use reciprocal interactions (retweet, quote, and reply) as a proxy of friendship, consistent with previous studies \cite{jiang2023retweet, rao2022partisan,ye2023susceptibility}. This filtering results in 49,794 and 73,960 users from the \textsc{Election} and \textsc{Covid} datasets, respectively. It is important to note that exposure to a user's tweet, as defined by unidirectional interactions, does not necessarily indicate friendship with that user. However, the reverse is true: if two users have mutual interactions and are considered friends, they are exposed to each other's tweets. Figure \ref{exposure_friend} illustrates the distinction between exposure and friendship.

\begin{figure*}[t]
  \centering
  \includegraphics[width=\textwidth]{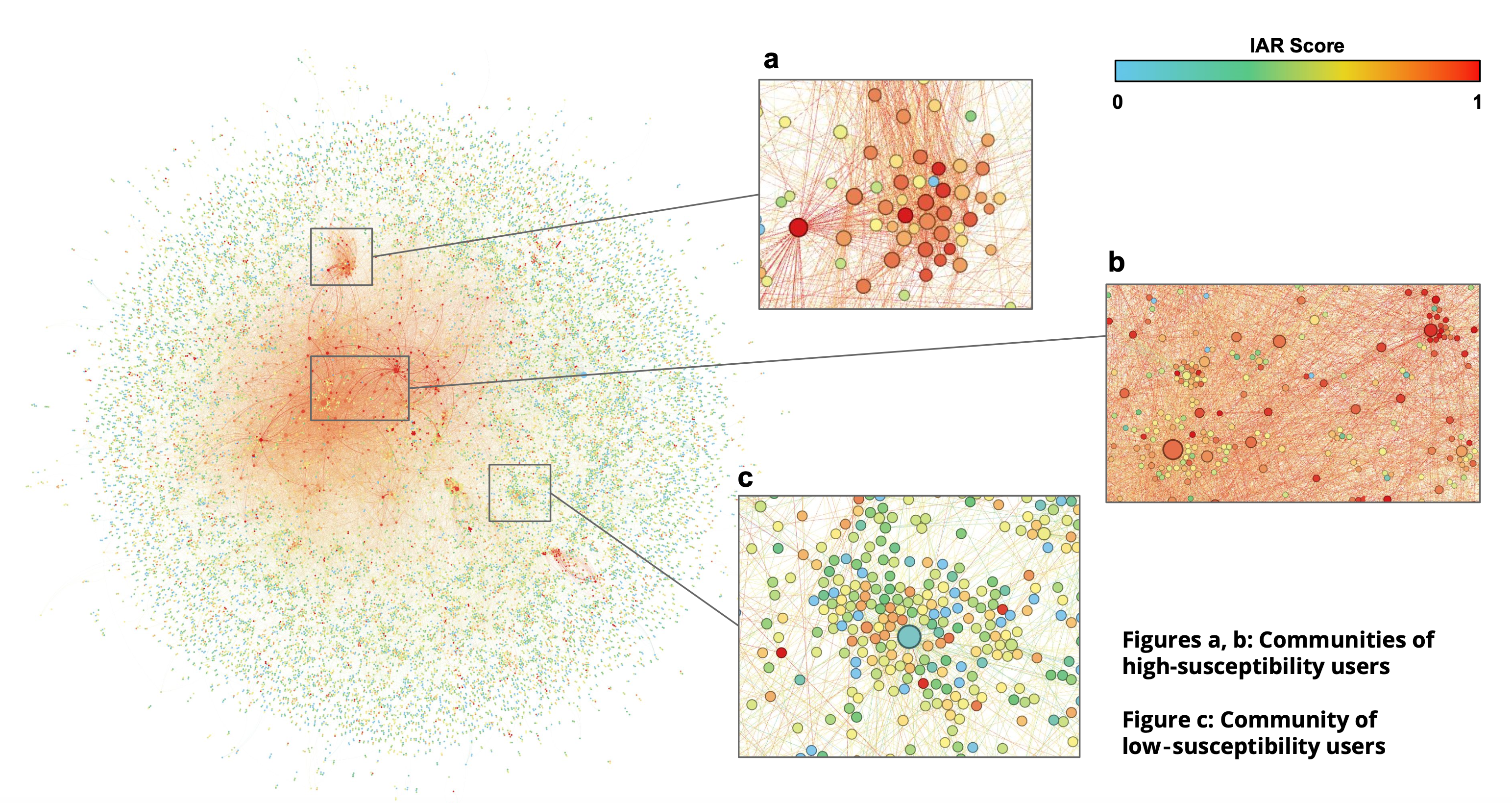}
  \caption{Homophily of influence‐driven adoption rates (IAR) in the \textit{Retweet} network of the \textsc{Election} dataset. Nodes are colored by IAR score (blue $=$ low susceptibility, red $=$ high susceptibility), scaled in size by node degree, and connected by edges representing mutual retweet interactions. Panels (a–c) display communities in which connected users share similar IAR scores, illustrating homophily.}
  \label{sus_visualization}
\end{figure*}


We construct three distinct friendship networks based on the types of interactions: \textit{Interaction}, \textit{Retweet}, and \textit{Mention} networks.
The \textit{Interaction} network includes all forms of interaction between users, such as retweets, quotes, and replies. In the \textit{Retweet} network, connections capture retweets only, whereas the \textit{Mention} network includes interactions through replies or quotes. We differentiate the \textit{Retweet} and \textit{Mention} networks due to their varying levels of social implications and endorsement, a distinction further explored in the ``Adoption" section. In Table \ref{dataset}, we present statistical descriptors of each network in the two datasets. 

\section{Results}

\subsection{RQ1: Susceptibility and Homophily}

We first examine whether susceptibility to influence exhibits homophily in a friendship network. To this end, we observe whether a user's susceptibility metrics are related to their friends.
Following \citet{jiang2023social}, we measure homophily by calculating the Pearson correlation between each user's susceptibility score and the weighted average susceptibility scores of all their friends. This method enables us to consider the frequency of user interactions and accounts for cases where users may have varying levels of influence across different connections. A coefficient of $-1$ suggests that users preferentially connect with others who have opposing susceptibility scores, while a coefficient of $1$ indicates a preference for connections between users with similar susceptibility scores.

\begin{figure}
    \centering
    \begin{subfigure}{.5\columnwidth}
        \centering        \includegraphics[width=\linewidth]{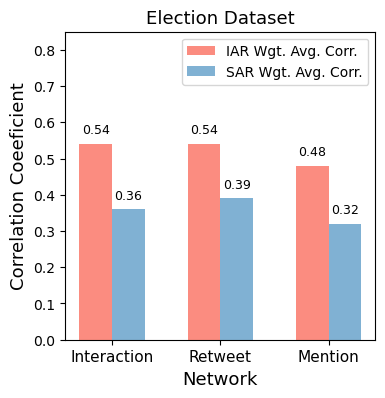}
        \label{WAC_election}
    \end{subfigure}%
    \begin{subfigure}{.5\columnwidth}
        \centering        \includegraphics[width=\linewidth]{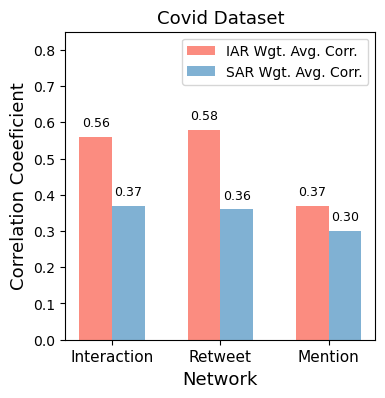}
        \label{WAC_covid}
    \end{subfigure}
    \vspace{-0.6cm}
    \caption{The susceptibility metrics (IAR and SAR) exhibit homophily within different friendship networks, as evidenced by the weighted average correlation (Wgt. Avg. Corr.) between users and their friends. All correlation coefficients are significant ($p < 0.001$).}
    \label{assort-WAC}
\end{figure}

The results presented in Figure \ref{assort-WAC} show that susceptibility is homophilous with respect to both the IAR and SAR scores, as indicated by the statistically significant and positive weighted average correlations. The results are consistent across all types of networks in both datasets. To further confirm the reliability of these findings, we compared our results with those from two null models. The first null model maintains the original network's degree distribution by randomly selecting and swapping the endpoints of edge pairs. The second null model preserves only the number of edges by randomly reassigning neighbors to nodes at one end of each edge. Both null models result in weighted average correlation measures that are not significant, therefore confirming that our results on the real data cannot be attributed to statistical flukes. The interested reader can refer to the \textit{Appendix} for further details.

\begin{figure*}[t]
    \centering    
    \includegraphics[width=12cm]{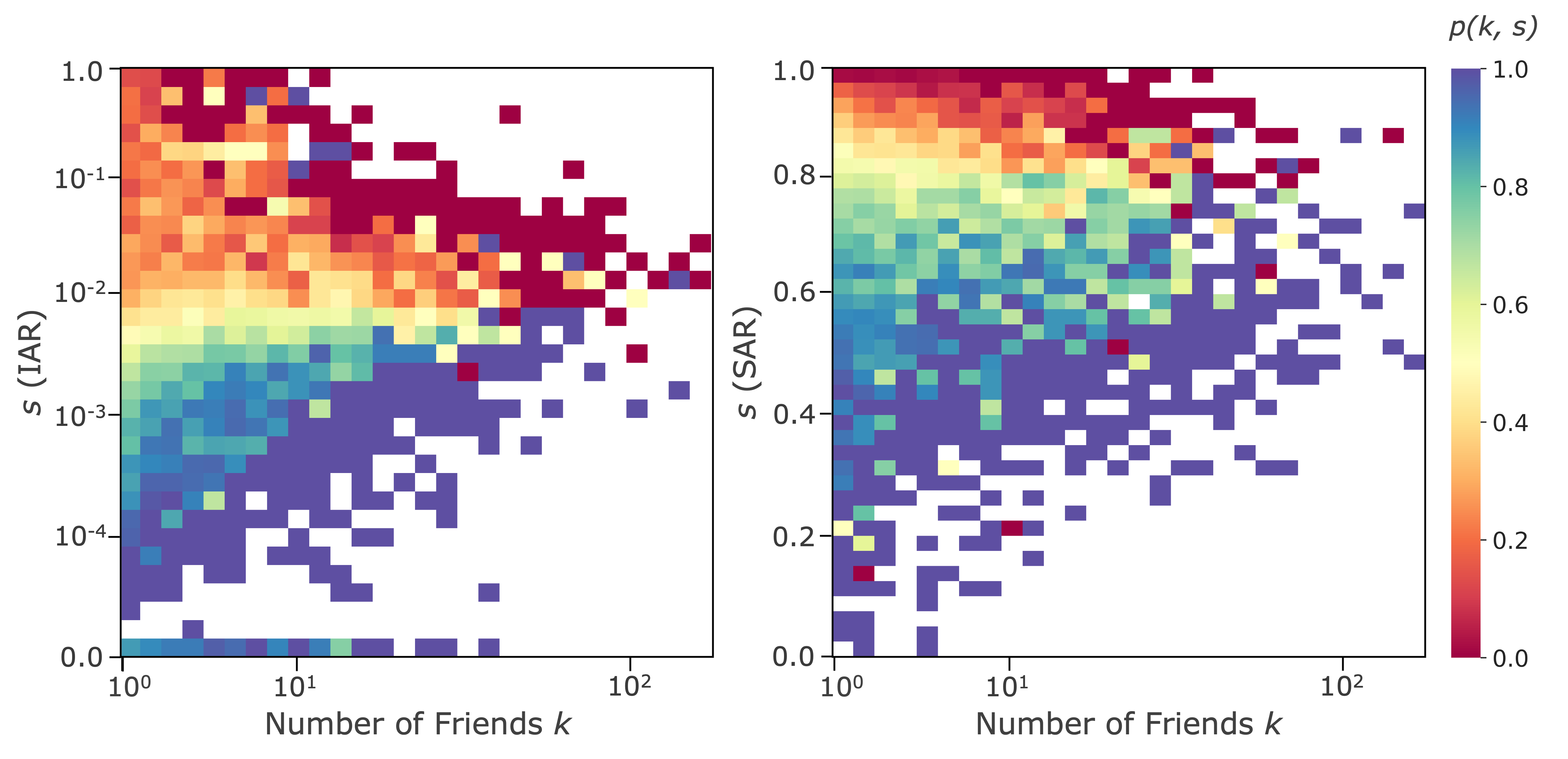} 
    \caption{Paradox holding probability $p(k, s)$ at a varying degree $k$ within the \textit{Interaction} network of the \textsc{COVID} dataset.}
    \label{IAR-SAR-heatmap}
\end{figure*}

Our analysis reveals that there is indeed homophily in user susceptibility. Moreover, this homophily, which refers to the tendency for individuals to form connections with others who are similar to them, appears to be more prevalent in influence-driven adoption (IAR) than in spontaneous adoption (SAR). Figure \ref{sus_visualization} visualizes the distribution of IAR scores in the \textit{Retweet} network of the \textsc{Election} dataset.
This suggests that within social networks, individuals connected by social ties tend to exhibit similar inclinations toward adopting the behaviors of their friends.
This is likely due to the stronger impact of peer influence within close social circles, where interactions are more frequent and intense \cite{de2010similarity}. As a result, a reinforcing loop emerges: individuals in tight-knit groups are more likely to adopt similar behaviors or attitudes due to the influence of their peers.


\paragraph{Summary.}
Susceptibility to influence is homophilous. In particular, influence-driven behavior (as measured by IAR) appears to be more homophilous than spontaneous-driven activity (as measured by SAR).


\subsection{RQ2: The \emph{Generalized Friendship Paradox} of Susceptibility to Influence}

When examining whether the \emph{Generalized Friendship Paradox} (GFP) applies to susceptibility to influence, we seek to determine whether individuals' friends are generally more susceptible to influence than they are. Consistent with \citet{eom2014generalized}, in this study, we formulate the GFP at both the \textit{individual} and \textit{network} levels. At the individual-level, a node exhibits GFP if its susceptibility metric is lower than the average of its friends. At the network-level, the GFP holds if the average susceptibility metric of all nodes is less than the average susceptibility metric of their respective friends.

\paragraph{Individual-level GFP.} 
The individual-level GFP holds for one user $i$ if the following condition is satisfied:
\begin{equation}
    s_i < \frac{\sum_{j \in \mathcal{N}_i} s_j}{k_i},
\end{equation}
where $s_i$ is the susceptibility metric of user $i$, $\mathcal{N}_i$ represents the set of $i$'s friends, and $k_i$ is the degree of $i$. Essentially, it calculates whether a user's susceptibility is less than the average of their friends. Based on this formulation, we define the paradox holding probability $p(k, s)$ as the proportion of nodes with degree $k$ and susceptibility metric $s$ that satisfies the condition in Eq. (1). 

\paragraph{Network-level GFP.} 
At the network-level, the GFP examines the phenomenon across the entire network. It is determined by comparing the average susceptibility metric of every node $\langle s \rangle$ with the average susceptibility metric of all their friends $\langle s \rangle_{nn}$, computed as follows: 
\begin{equation}
    \langle s \rangle_{nn} = \frac{\sum_{i=1}^N k_i s_i}{\sum_{i=1}^N k_i},
\end{equation}
where a node $i$ with degree $k_i$ is counted as a neighbor $k_i$ times.
If the average susceptibility metric of a node's friends exceeds the node's own average susceptibility metric ($\langle s \rangle < \langle s \rangle_{nn}$),
then the GFP holds at the network-level. 

\paragraph{GFP: Comparing SAR vs. IAR.}
Figure \ref{IAR-SAR-heatmap} depicts the paradox holding probability $p(k, s)$ as a function of degree $k$ and susceptibility metric $s$ using the \textit{Interaction} network in the \textsc{Covid} dataset. Similar findings from the \textsc{Election} dataset are presented in the \textit{Appendix}. 
This three-dimensional representation reveals that, with a constant degree $k$, $p(k, s)$ decreases as $s$ (IAR or SAR) increases, aligning with findings from \citet{eom2014generalized}. This implies that users with very low susceptibility $s$ tend to connect with those having higher $s$ values, resulting in a $p(k, s)$ close to $1$. Conversely, users with very high susceptibility values exhibit a $p(k, s)$ approaching $0$. Users across this spectrum exhibit a consistent trend, wherein $p(k, s)$ remains close to 1 when $s$ and $k$ are strongly, positively correlated. This effect is particularly evident for IAR, indicating a specific pattern where the paradox holds.

\begin{table*}
  \centering
  \small
  \begin{tabular}{@{}lcrcccccccc@{}}
  \toprule
    \multirow{3}{*}{\textbf{Network}} & \multirow{3}{*}{\textbf{Metric} $s$} & \multirow{3}{*}{\textbf{$\rho_{ks}$}} & \multicolumn{3}{c}{Individual-level GFP} & \multicolumn{5}{c}{Network-level GFP} \\
    \cmidrule(lr){4-6} \cmidrule(lr){7-11}
     & & & 
        \begin{tabular}{@{}c@{}} \textbf{Real-Network} \\ $P$ \end{tabular} & 
        \begin{tabular}{@{}c@{}} \textbf{Baseline1} \\ $P$ \end{tabular} & 
        \begin{tabular}{@{}c@{}} \textbf{Baseline2} \\ $P$ \end{tabular} & 
    \textbf{$\langle s \rangle$} & & 
        \begin{tabular}{@{}c@{}} \textbf{Real-Network} \\ $\langle s \rangle_{nn}$ \end{tabular} & 
        \begin{tabular}{@{}c@{}} \textbf{Baseline1} \\ $\langle s \rangle_{nn}$ \end{tabular} & 
        \begin{tabular}{@{}c@{}} \textbf{Baseline2} \\ $\langle s \rangle_{nn}$ \end{tabular} \\
    \midrule
    \multirow{2}{*}{Interaction} & IAR & 0.179 & 0.541 & 0.672 & 0.798 & 0.007 & $<$ & 0.008 & 0.019 & 0.028\\
    & SAR & $-$0.166 & 0.409 & 0.326 & 0.249 & 0.880 & $>$ & 0.868 & 0.810 & 0.751\\
    \midrule
    \multirow{2}{*}{Retweet} & IAR & 0.298 & 0.501 & 0.662 & 0.824 & 0.013 & $<$ & 0.014 & 0.022 & 0.040\\
    & SAR & $-$0.294 & 0.425 & 0.355 & 0.250 & 0.821 & $>$ & 0.809 & 0.722 & 0.670\\
    \midrule
    \multirow{2}{*}{Mention} & IAR & 0.111 & 0.544 & 0.623 & 0.697 & 0.005 & $<$ & 0.006 & 0.009 & 0.015\\
    & SAR & $-$0.099 & 0.383 & 0.354 & 0.325 & 0.896 & $>$ & 0.880 & 0.852 & 0.834\\
    \bottomrule
\end{tabular}
  \caption{Empirical results of the \emph{Generalized Friendship Paradox} (GFP) in the three friendship networks of the \textsc{Election} dataset. For each susceptibility metric $s$ (both IAR and SAR), we calculate the Spearman correlation coefficient $\rho_{ks}$ between degree $k$ and $s$ ($p < 0.001$ for all $\rho_{ks}$), average paradox holding probability $P$, and the average susceptibility $\langle s \rangle$ of nodes and their friends $\langle s \rangle_{nn}$.}
\label{GFP}
\end{table*}

Table \ref{GFP} presents the empirical results 
of the individual- and network-level GFP across three friendship networks constructed from the \textsc{Election} dataset. Similar results are observed in the \textsc{Covid} dataset.
For each susceptibility metric $s$ (IAR or SAR), we calculate the average paradox holding probability at the individual-level $P$, integrating $p(k,s)$ over $k$ and $s$, following \citet{eom2014generalized}, and the Spearman correlation coefficient between node degree $k$ and susceptibility metric $s$, denoted as $\rho_{ks}$. For the network-level GFP, we compare the node average $\langle s \rangle$ and their friends' average $\langle s \rangle_{nn}$. We also employ two baseline models for comparison based on the null models described above.

For IAR, 
the value of $P$ exceeds 0.5 in every friendship network, indicating that the GFP holds at the individual-level. 
Furthermore, the condition $\langle s \rangle < \langle s \rangle_{nn}$ is consistently satisfied across networks, indicating that the GFP also holds at the network-level. This implies that, on average, individuals' friends are more likely to adopt content they are exposed to than the individuals themselves. In contrast, for SAR, $P$ consistently falls below 0.5, and $\langle s \rangle$ is always higher than $\langle s \rangle_{nn}$. These observations suggest that the GFP does not hold for spontaneous adoption. In simpler terms, individuals are less inclined to adopt unseen content compared to their friends.

Moreover, it is worth noting that while the GFP holds for IAR, it is less pronounced compared to the two null models. The discrepancy between individual IAR and that of their friends is smaller in the observed network than in a randomly constructed network. For example, at the individual-level, in the \textit{Interaction} network of the \textsc{Election} dataset, a random baseline model predicts that in 79.8\% of cases, an individual's friends, on average, would exhibit higher IAR than the individual. However, the actual observed rate in the real-world network is only 54.1\%. At the network-level, the differences between $\langle s \rangle$ and $\langle s \rangle_{nn}$ are also smaller in the actual network. This  discrepancy between random baselines and actual networks is similarly observed for SAR.

These observations corroborate the study by \citet{lerman2016majority}, which demonstrates that the ``majority illusion,'' a phenomenon resulting from the friendship paradox, is more intense within networks characterized by heterogeneous degree distributions and disassortative structures. Our observations of real-world social media networks, which exhibit greater homophily compared to random baselines, indeed reveal a reduced paradox holding probability. This implies that the structural attributes of a network can significantly impact the manifestation of the GFP.

\paragraph{Summary.}
Our findings indicate that the \emph{Generalized Friendship Paradox} holds for influence-driven behavior but not for spontaneous actions on social networks. This suggests that users' friends are generally more susceptible to influence and less prone to spontaneous sharing activities than the users themselves. 



\subsection{RQ3: Predicting Susceptibility to Online Influence}
As our findings in RQ1 and RQ2 provide valuable insights into the relationship between an individual's susceptibility and that of their friends, we explore the potential to predict a user's susceptibility based on the susceptibility levels of their friends. Furthermore, we evaluate whether incorporating additional features, such as user and friend metadata along with network properties, can enhance prediction accuracy. To accomplish this, we employ both simple linear regression and random forest models.

\paragraph{Linear Regression Model.}
First, we assess the effectiveness of predicting a user's IAR (resp. SAR) based on the average IAR (resp. SAR) of their friends. The linear regression model is expressed as:
\begin{equation}
s_{user} = \beta_0 + \beta_1 \times \bar{s}_{friend} + \epsilon,
\end{equation}
where $s_{user}$ and $\bar{s}_{friend}$ refer to a user IAR (resp. SAR) and their friends' average IAR (resp. SAR). 

Table \ref{linear_regression} shows the results of the simple linear regression analysis. From Table \ref{linear_regression}, we note that both IAR and SAR models yield positive regression coefficients ($\beta_1$), suggesting a positive relationship between a user's susceptibility metrics and those of their friends. Notably, the $R^2$ values for the IAR models are consistently higher than those of the SAR models, indicating that influence-driven adoptions are more accurately explained by friends' susceptibility compared to spontaneous adoptions. 

Additionally, we also attempt to predict user IAR using friends' SAR and vice versa; however, these models produce $R^2$ values all below 0.05 in both datasets, highlighting that although strongly negative-correlated, IAR and SAR do not adequately predict each other.

\begin{table}
  \centering
  \small
  \begin{tabular}{lccccc}
  \toprule
    Dataset & $s_{user}$ & $\bar{s}_{friend}$ & $\beta_0$ & $\beta_1$ & $R^2$ \\
    \midrule
    \textsc{Election} & IAR & IAR & 0.002 & 0.622 & 0.363\\
    \textsc{Election} & SAR & SAR & 0.550 & 0.377 & 0.137\\
    \textsc{Covid} & IAR & IAR & 0.005 & 0.533 & 0.283\\
    \textsc{Covid} & SAR & SAR & 0.537 & 0.392 & 0.134\\
    \bottomrule
\end{tabular}
  \caption{Summary of the linear regression results for predicting user susceptibility metrics based on the friends' average susceptibility metrics. All regression coefficients reported are significant ($p < 0.001$).}
  \label{linear_regression}
\end{table}

\paragraph{Random Forest Model.}
To assess whether a more advanced model equipped with a broader range of features can enhance predictive accuracy, we employ an interpretable ensembling machine learning algorithm. This approach also aims to enhance our understanding of the factors that impact users' susceptibility to social influence. Specifically, 
we utilize a random forest regression model that incorporates not only susceptibility metrics but also network properties and user metadata. This feature set includes friends count, followers count, favorites count (the total number of tweets a user has liked), and statuses count (the total number of tweets a user has posted), along with network properties from the \textit{Interaction} network such as degree centrality, eigenvector centrality, and clustering coefficient.


We divide users into training and testing sets and use the variance inflation factor (VIF) to assess multicollinearity among the features. In both the \textsc{Election} and \textsc{Covid} datasets, all VIF values are below 5, indicating no significant multicollinearity issues \cite{james2013introduction}. We optimize the random forest regressor using a randomized search over the hyperparameter space to efficiently predict user IAR and SAR scores. 
The randomized search is configured to sample 100 different parameter settings using 5-fold cross-validation. After selecting the optimal parameters (see the \textit{Appendix} for details), we assess each model's performance on the test set using the $R^2$ metric.

Table \ref{random_forest} provides an overview of the results. The random forest regressor shows promising performance, particularly for the IAR metric. It slightly outperforms the linear regressor, as indicated by the $R^2$ difference labeled as $\Delta R^2$ in Table \ref{random_forest}. This difference is more pronounced for the SAR metric, while it remains quite small (less than 0.05) for the IAR metric, suggesting that IAR can be effectively predicted using solely friends' IAR, as previously demonstrated in the linear regression model. An analysis of feature importance further supports this intuition.

\begin{table}
  \centering
  \small
  \begin{tabular}{lcccc}
  \toprule
    & \multicolumn{2}{c}{\textsc{Election}} & \multicolumn{2}{c}{\textsc{Covid}}\\
    \cmidrule(lr){2-3} \cmidrule(lr){4-5}
    \textit{Susceptibility Metric} & IAR & SAR & IAR & SAR\\
    \midrule
    Random Forest $R^2$ & 0.406 & 0.389 & 0.332 & 0.312\\
    Linear Regression $R^2$ & 0.363 & 0.137 & 0.283 & 0.134\\
    $\Delta R^2$ & 0.043 & 0.252 & 0.049 & 0.178\\
    \bottomrule
\end{tabular}
  \caption{Summary of the performance of the random forest and linear regressors. $R^2$ denotes the coefficient of determination and $\Delta R^2$ represents the $R^2$ difference between random forest and linear regression models.}
  \label{random_forest}
\end{table}

\begin{figure}
    \centering    
    \begin{subfigure}{.49\columnwidth}
        \centering        \includegraphics[width=\linewidth]{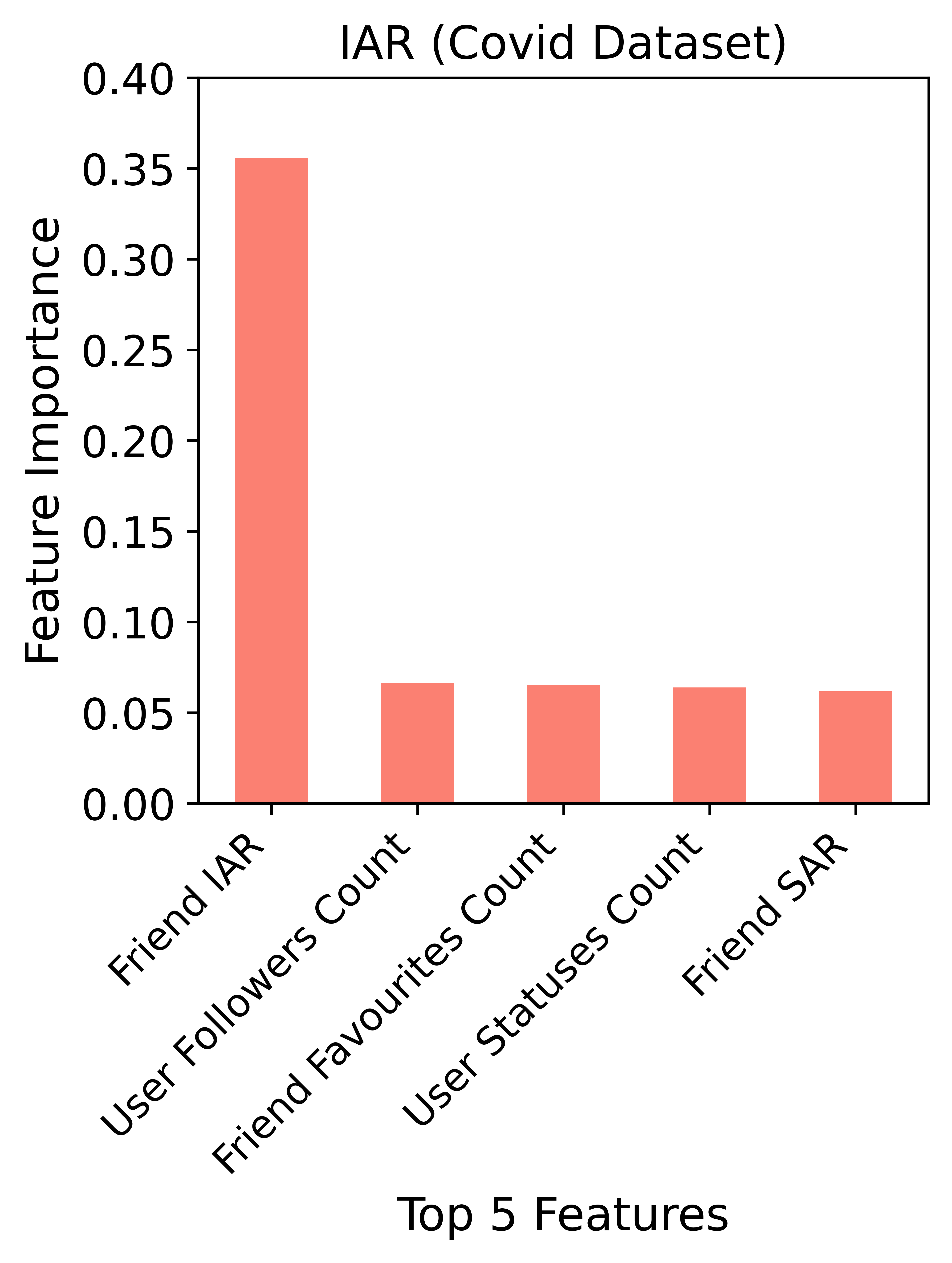}        \label{IAR_feature_importance_covid}
    \end{subfigure}%
    \begin{subfigure}{.49\columnwidth}
        \centering        \includegraphics[width=\linewidth]{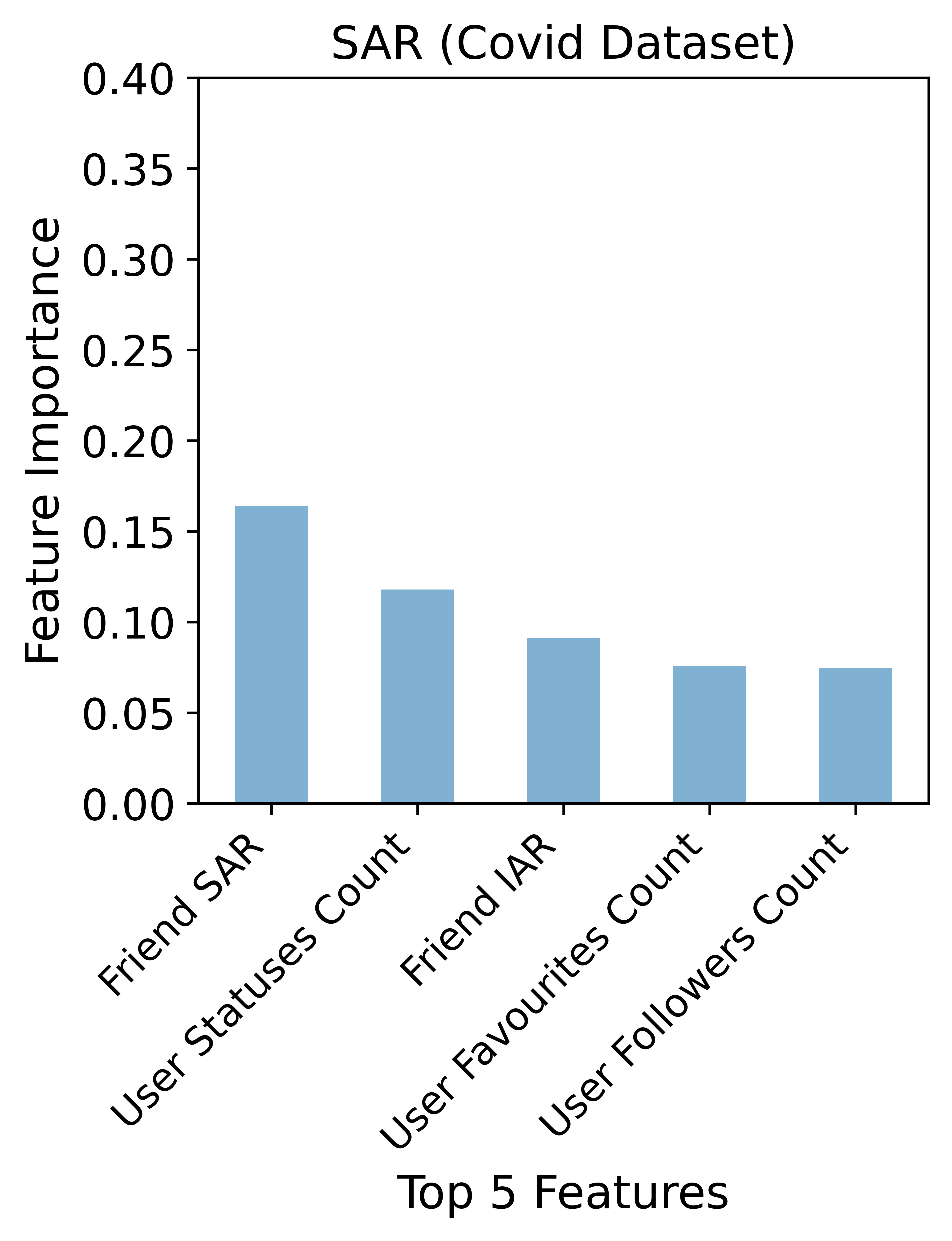}        \label{SAR_feature_importance_covid}
    \end{subfigure}
    \vspace{-0.6cm}
    \caption{The top 5 most important features from the random forest model in the \textsc{COVID} dataset for predicting IAR and SAR.}
    \label{feature_imp}
    \vspace{-0.2cm}
\end{figure}


Figure \ref{feature_imp} illustrates the feature importance (top 5 features) for the random forest regressor in predicting IAR and SAR within the \textsc{Covid} dataset. Notably, \textit{friends' IAR} emerges as the most significant feature for predicting IAR, with other features proving substantially less relevant. This finding confirms our previous intuition and remains consistent in the \textsc{Election} dataset.
For SAR predictions, while \textit{friends' SAR} is the most critical feature, additional factors such as \textit{statuses count, friends' IAR}, and \textit{favorites count} also contribute. This pattern holds in the \textsc{Election} dataset, where \textit{statuses count} is the most important feature.

\begin{figure*}
    \centering
    \begin{minipage}{0.48\textwidth}
        \centering
        \includegraphics[width=\textwidth]{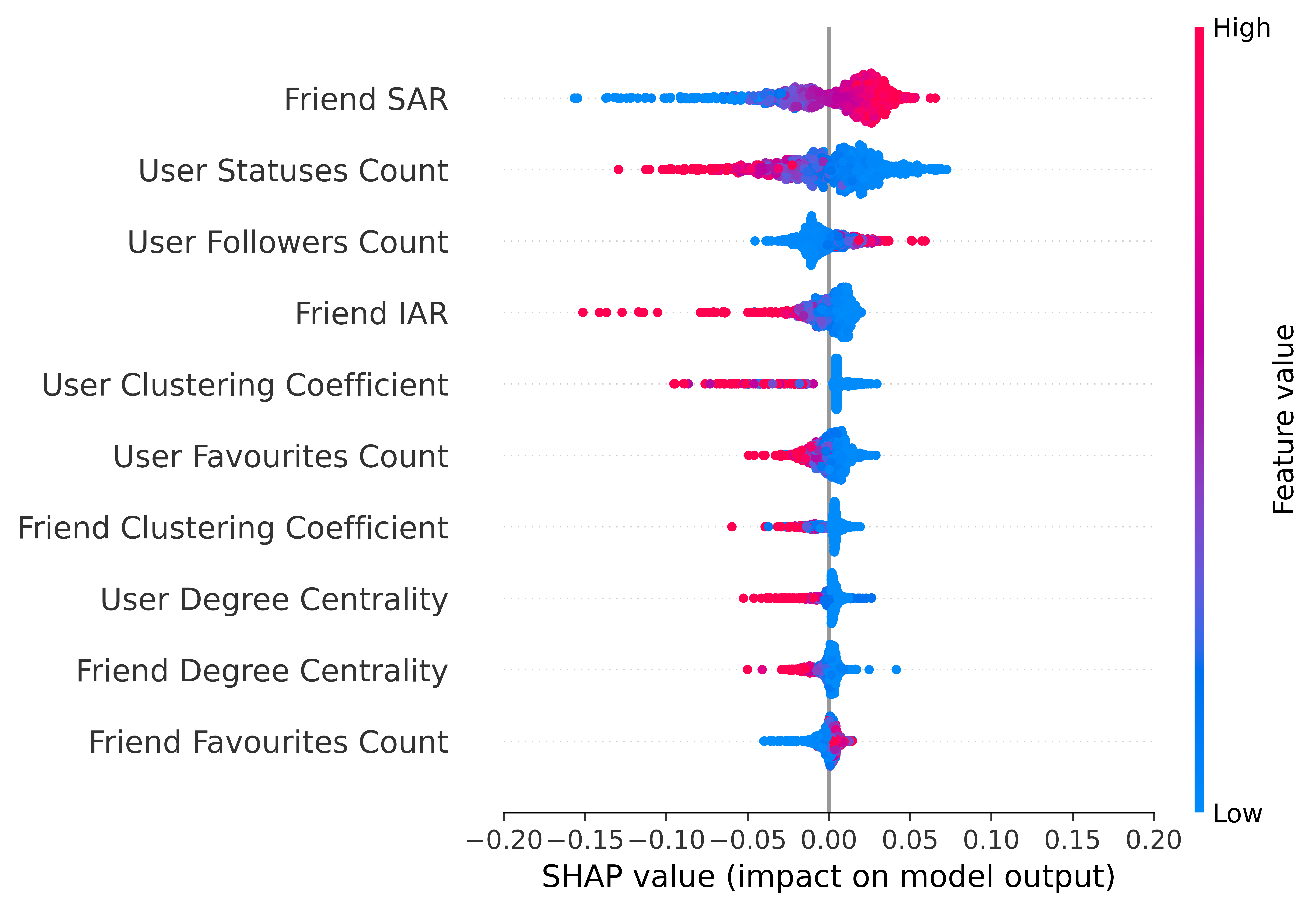}
        \caption*{(a) SHAP value distributions for SAR predictions.}
    \end{minipage}%
    \hfill
    \begin{minipage}{0.48\textwidth}
        \centering
        \includegraphics[width=\textwidth]{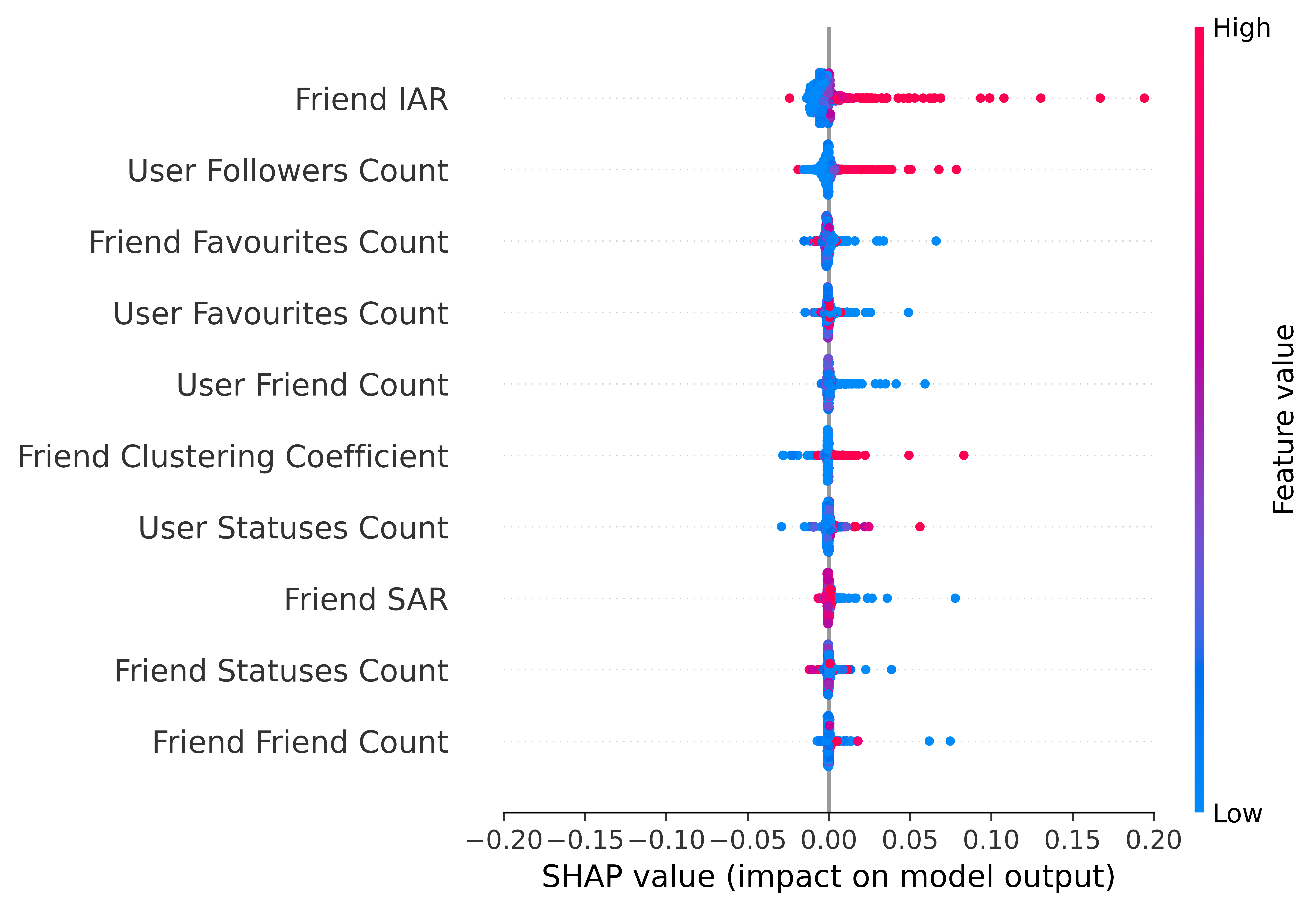}
        \caption*{(b) SHAP value distributions for IAR predictions.}
    \end{minipage}
    \caption{SHAP value distributions for SAR (a) and IAR (b) predictions in the \textsc{Covid} dataset. Each distribution illustrates the direction and magnitude of the impact of individual features on the model's output, with features ordered by importance from top to bottom.}
    \label{SHAP-covid-horizontal}
\end{figure*}

\paragraph{Output Explainability.} To further elucidate the results from the feature importance analysis and quantify the specific contribution of each feature to individual predictions, we compute the SHAP (SHapley Additive exPlanations) values \cite{lundberg2017unified} of the dependent variables predicted in the test set. SHAP values provide a nuanced measure of the impact of each feature on individual instances. Each feature is assigned a SHAP value per instance, which can be aggregated to offer a comprehensive view of the direction and magnitude of its impact on the model's predictions. Figure \ref{SHAP-covid-horizontal} shows a SHAP summary plot for the SAR and IAR metrics of the \textsc{Covid} dataset (additional results are detailed in the \textit{Appendix}). On the y-axis, features are listed in order of impact, while the x-axis displays the SHAP values, indicating the change in log odds caused by each feature. Points represent individual data instances, colored red (resp. blue) for high (resp. low) feature values.

As shown by Figure \ref{SHAP-covid-horizontal}, the most impactful features affecting the SAR predictions in the \textsc{Covid} dataset are \textit{friends' SAR, user statuses count}, and \textit{user clustering coefficient}. The features \textit{friends' SAR} and \textit{user statuses count} have significant opposing impacts on the predictions, indicating that these features can notably increase or decrease the SAR value. Specifically, high values of \textit{friends' SAR} tend to increase the SAR score, while higher values of \textit{user statuses count} tend to decrease it. This relationship between \textit{user statuses count} and SAR suggests that a high volume of shared tweets correlates with a small rate of spontaneous adoptions.
Interestingly, the \textit{user clustering coefficient} also emerges as a significant factor, where higher clustering coefficients typically decrease SAR, possibly indicating that users with more cohesive social networks may experience fewer spontaneous adoptions. This finding is consistent in the \textsc{Election} dataset (see the \textit{Appendix}), where user statuses count, friends' SAR, and user favorites count are the most impactful features. 

For IAR prediction (see Figure \ref{SHAP-covid-horizontal}), the most impactful feature is \textit{friends' IAR}, with high values of \textit{friends' IAR} significantly increasing the IAR score, supporting our previous results related to homophily and GFP in IAR.  

\paragraph{Summary.} Users' level of influence-driven adoption can be effectively predicted by solely leveraging friends' IAR. In contrast, the predictability of a user's SAR is significantly enhanced by incorporating additional features, such as counts of tweets and likes. Explainable models reveal that SAR is affected by a wider array of factors, suggesting a more complex interplay of user engagement and characteristics in the spontaneous adoption of online content. 

\subsection{Sensitivity Analysis}

To address potential sampling biases related to user network construction, we conducted a sensitivity analysis by testing different thresholds for identifying target users. Specifically, we applied thresholds of 5 and 20 shared URLs, in addition to the previously used threshold of 10 URLs, to select the set of target users under scrutiny. This approach allowed us to assess whether varying levels of user activity affected our results. 
The results, detailed in the \textit{Appendix}, remained consistent across the different thresholds. Both the IAR and SAR metrics exhibit homophily. Additionally, the \emph{Generalized Friendship Paradox} (GFP) holds for influence-driven behaviors (IAR) but not for spontaneous actions (SAR), further validating our findings and conclusions.

\section{Discussion and Conclusions}

This study, based on the analysis of millions of tweets, explores the networked dynamics of user susceptibility to online influence from the dual perspective of influence-driven and spontaneous adoption. We present four key insights drawn from our research.

\paragraph{Finding 1: Susceptibility is homophilous.}
We find that susceptibility is homophilous in social networks, as evidenced by the positive correlations between users' adoption rates (either influence-driven or spontaneous) and those of their friends. This indicates that influenceable individuals often connect with similarly susceptible peers. 
Spontaneous adoption shows significant but lower homophily.

\paragraph{Finding 2a: The Susceptibility Paradox holds for influence-driven behavior.}
We observe the \emph{Generalized Friendship Paradox} (GFP) for user susceptibility to influence, termed \textit{the Susceptibility Paradox}, across two dimensions: First, individuals in a social network tend to be connected to others who are more likely to adopt content due to prior exposure, as reflected in a higher IAR. Second, users' friends engage in less spontaneous content adoption with respect to the users themselves, as indicated by their lower SAR. Collectively, these observations suggest that ``your friends are generally more susceptible to social influence than you.''
Additionally, we observe that IAR and SAR are negatively correlated, suggesting that users with higher influence-driven adoption are less likely to share content spontaneously. 
%

\paragraph{Finding 2b: The Susceptibility Paradox is less pronounced when homophily is present.} While we observe that the GFP holds for IAR but not for SAR, 
both IAR and SAR demonstrate homophily. These homophilous patterns indicate that individuals are prone to connect with others who exhibit similar rates of influence-driven and spontaneous adoption. This reveals two countervailing forces: while an individual's friends generally have a higher IAR but a lower SAR compared to the individual, their overall susceptibility scores tend to remain similar.
%
Comparing the individual- and network-level GFP in homophilous real-world networks against non-homophilous baseline networks, our findings suggest that the GFP is less pronounced in homophilous networks, where high-IAR (or SAR) nodes tend to connect with other high-IAR (or SAR) nodes. 
This corroborates the findings of \citet{lerman2016majority}, which indicate that the ``majority illusion" is more pronounced in heterogeneous, disassortative networks.


\paragraph{Finding 3: IAR can be predicted by friends' IAR and is more predictable than SAR.} Our analysis indicates that IAR can be predicted solely based on friends' IAR, whereas SAR is influenced by a broader range of factors. This distinction adds depth to our understanding of susceptibility to influence. On the one hand, our results suggest that influence-driven behaviors (IAR) tend to follow more predictable patterns. This likely stems from the strong relationship between influence-driven adoption and topological properties, such as homophily (RQ1) and GFP (RQ2), observed in this study. Also, research has shown that targeted marketing and influence campaigns can be effectively strategized based on previous adoption patterns of a consumer's network \cite{hill2006network}. On the other hand, spontaneous behaviors (SAR) are influenced by a myriad of internal and external, often stochastic factors, making them less predictable.

\paragraph{Implications.}
Our research offers practical insights for social media platforms, researchers, policymakers, and marketers.
The observed homophily in susceptibility suggests that targeted interventions in fields like public health and marketing can effectively target influenceable audiences. By identifying these cohesive, vulnerable groups, intervention strategies can be tailored to mitigate the spread of harmful content and encourage positive behaviors more efficiently \cite{centola2011experimental}. Further, the observation that the \emph{Generalized Friendship Paradox} applies to influence-driven adoption could inspire efficient sampling algorithms to identify highly susceptible populations \cite{alipourfard2020friendship}. 

Predicting users' IAR and SAR could be valuable in two key scenarios. First, for new users joining an existing social network, their IAR and SAR can be approximated based on interactions with more longevous users, leveraging metrics and attributes of their friends. This allows for an early assessment of a new user’s susceptibility to influence. Second, for less active users or those who rarely share URLs---falling below the threshold required to compute their IAR or SAR---we can infer their susceptibility leveraging the susceptibility metrics of their friends. Overall, this enables social media providers to account for the susceptibility of users with limited activity. Since IAR is more predictable, and individuals with high IAR are more influenced by their social connections, platforms can strategically target vulnerable audiences by nudging them with accurate, fact-checked content to help prevent the endorsement and spread of misinformation \cite{pennycook2021shifting}.

Future research will extend our susceptibility framework across various media and content modalities. Since the framework is based solely on the dynamics between exposure and adoption, it is platform-agnostic, allowing us to analyze susceptibility on different social media networks. This could involve studying not only URLs in general, but also disentangling specific content types, such as videos, images, and memes. Additionally, our research will explore application scenarios where predicting user susceptibility is critical, such as modeling the impact of coordinated influence operations \cite{luceri2024unmasking} and exposure to AI-generated disinformation \cite{augenstein2024factuality} on content adoption. Validating our model in these domains will further strengthen its real-world applicability and relevance.

\paragraph{Limitations.}
We acknowledge several limitations in our research. First, our study draws on datasets specifically related to the 2020 US election and the COVID-19 pandemic, capturing only a subset of users over an extended but limited period. Consequently, generalizing our results to broader user networks, different contexts, or other social media platforms necessitates further validation. 
Second, we infer exposure based on observable user interactions, which might not comprehensively capture exposures within Twitter and across different media. Nonetheless, this method for assessing exposure was tested and validated in prior research \cite{ferrara2015measuring, rao2023retweets, ye2023susceptibility}. 
Third, our network construction method excludes less active users (those sharing fewer than 5, 10, or 20 URLs), which may limit the generalizability of the observed homophily and friendship paradox. Additionally, since we use URL sharing as the adoption criterion, there may be users who rarely share URLs, further restricting the size of the target users under scrutiny. However, our sensitivity analysis shows consistent findings (see related section in the \textit{Appendix}). Varying the target user thresholds (sharing 5, 10, or 20 URLs) did not affect the significance of our results, indicating the robustness of our findings despite potential selection bias.
Fourth, although friends' IAR explains 36\% and 28\% of the variation of user IAR in \textsc{Election} and \textsc{Covid} datasets, respectively, we recognize that social influence is a complex phenomenon and may be influenced by various unobserved personal, societal, and cultural factors. Despite these limitations, the consistency of our results across multiple intertwined analyses on distinct datasets supports the reliability and robustness of our findings.

\section{Acknowledgments}
Work supported in part by DARPA (contract \#HR001121C0169) and the NSF (award \#2331722).

\bibliography{aaai22}

\newpage
\section{Paper Checklist}

\begin{enumerate}

\item General questions:
\begin{itemize}
    \item [(a)] \textbf{Would answering this research question advance science without violating social contracts, such as violating privacy norms, perpetuating unfair profiling, exacerbating the socio-economic divide, or implying disrespect to societies or cultures?}
        \begin{itemize}
            \item \answerYes{Yes, the study adheres to ethical guidelines, focusing on aggregate data analysis without violating individual privacy.}
        \end{itemize}
    \item [(b)] \textbf{Do your main claims in the abstract and introduction accurately reflect the paper’s contributions and scope?}
        \begin{itemize}
            \item \answerYes{Yes, the main claims regarding the Susceptibility Paradox in influence on social media are supported by the findings.}
        \end{itemize}
    \item [(c)] \textbf{Do you clarify how the proposed methodological approach is appropriate for the claims made?}
        \begin{itemize}
            \item \answerYes{Yes, the methodological approach, including statistical analysis and machine learning models, is detailed and appropriate for the research questions.}
        \end{itemize}
    \item [(d)] \textbf{Do you clarify what are possible artifacts in the data used, given population-specific distributions?}
        \begin{itemize}
            \item \answerYes{Yes, the paper discusses potential limitations related to data collection and the representativeness of the sample populations.}
        \end{itemize}
    \item [(e)] \textbf{Did you describe the limitations of your work?}
        \begin{itemize}
            \item \answerYes{Yes, the manuscript includes a section on limitations, acknowledging the constraints of the data and methods used.}
        \end{itemize}
    \item [(f)] \textbf{Did you discuss any potential negative societal impacts of your work?}
        \begin{itemize}
            \item \answerYes{Yes, the potential societal impacts, particularly in terms of misuse for targeted manipulation or misinformation, are discussed.}
        \end{itemize}
    \item [(g)] \textbf{Did you discuss any potential misuse of your work?}
        \begin{itemize}
            \item \answerYes{Yes, the manuscript addresses potential misuse, especially regarding the exploitation of susceptibility metrics for unethical influence campaigns.}
        \end{itemize}
    \item [(h)] \textbf{Did you describe steps taken to prevent or mitigate potential negative outcomes of the research, such as data and model documentation, data anonymization, responsible release, access control, and the reproducibility of findings?}
        \begin{itemize}
            \item \answerYes{Yes, measures such as data anonymization and ethical guidelines for data usage are mentioned.}
        \end{itemize}
    \item [(i)] \textbf{Have you read the ethics review guidelines and ensured that your paper conforms to them?}
        \begin{itemize}
            \item \answerYes{Yes, the authors confirm adherence to ethics review guidelines throughout the study.}
        \end{itemize}
\end{itemize}

\item Hypotheses testing:
\begin{itemize}
    \item [(a)] \textbf{Did you clearly state the assumptions underlying all theoretical results?}
        \begin{itemize}
            \item \answerYes{Yes, the assumptions underlying the statistical models and hypotheses are clearly stated.}
        \end{itemize}
    \item [(b)] \textbf{Have you provided justifications for all theoretical results?}
        \begin{itemize}
            \item \answerYes{Yes, the theoretical results are justified with empirical data and statistical analysis.}
        \end{itemize}
    \item [(c)] \textbf{Did you discuss competing hypotheses or theories that might challenge or complement your theoretical results?}
        \begin{itemize}
            \item \answerYes{Yes, the paper discusses alternative explanations and competing theories regarding social influence and susceptibility.}
        \end{itemize}
    \item [(d)] \textbf{Have you considered alternative mechanisms or explanations that might account for the same outcomes observed in your study?}
        \begin{itemize}
            \item \answerYes{Yes, alternative mechanisms and explanations are explored to ensure the robustness of the findings.}
        \end{itemize}
    \item [(e)] \textbf{Did you address potential biases or limitations in your theoretical framework?}
        \begin{itemize}
            \item \answerYes{Yes, the potential biases and limitations in the theoretical framework are acknowledged and discussed.}
        \end{itemize}
    \item [(f)] \textbf{Have you related your theoretical results to the existing literature in social science?}
        \begin{itemize}
            \item \answerYes{Yes, the results are related to existing social science literature on social influence and susceptibility.}
        \end{itemize}
    \item [(g)] \textbf{Did you discuss the implications of your theoretical results for policy, practice, or further research in the social science domain?}
        \begin{itemize}
            \item \answerYes{Yes, the implications for policy, practice, and further research are thoroughly discussed.}
        \end{itemize}
\end{itemize}

\item Theoretical proofs:
\begin{itemize}
    \item [(a)] \textbf{Did you state the full set of assumptions of all theoretical results?}
        \begin{itemize}
            \item \answerYes{Yes, all assumptions for the theoretical results are stated clearly.}
        \end{itemize}
    \item [(b)] \textbf{Did you include complete proofs of all theoretical results?}
        \begin{itemize}
            \item \answerYes{Yes, mathematical formulations are included where applicable.}
        \end{itemize}
\end{itemize}

\item Machine learning experiments:
\begin{itemize}
    \item [(a)] \textbf{Did you include the code, data, and instructions needed to reproduce the main experimental results (either in the supplemental material or as a URL)?}
        \begin{itemize}
            \item \answerYes{Yes, details on code and data will be provided to ensure reproducibility.}
        \end{itemize}
    \item [(b)] \textbf{Did you specify all the training details (e.g., data splits, hyperparameters, how they were chosen)?}
        \begin{itemize}
            \item \answerYes{Yes, the training details, including data splits and hyperparameters, are specified.}
        \end{itemize}
    \item [(c)] \textbf{Did you report error bars (e.g., with respect to the random seed after running experiments multiple times)?}
        \begin{itemize}
            \item \answerYes{Yes, error bars and statistical significance are reported where applicable.}
        \end{itemize}
    \item [(d)] \textbf{Did you include the total amount of compute and the type of resources used (e.g., type of GPUs, internal cluster, or cloud provider)?}
        \begin{itemize}
            \item \answerNA{NA}
        \end{itemize}
    \item [(e)] \textbf{Do you justify how the proposed evaluation is sufficient and appropriate to the claims made?}
        \begin{itemize}
            \item \answerYes{Yes, the evaluation methods are justified as appropriate for the research claims.}
        \end{itemize}
    \item [(f)] \textbf{Do you discuss what is ``the cost'' of misclassification and fault (in)tolerance?}
        \begin{itemize}
            \item \answerYes{Yes, the costs of misclassification are prediction inaccuracy are discussed in the context of the study.}
        \end{itemize}
\end{itemize}

\item Code \& Data Assets:
\begin{itemize}
    \item [(a)] \textbf{If your work uses existing assets, did you cite the creators?}
        \begin{itemize}
            \item \answerYes{Yes, existing assets and their creators are properly cited.}
        \end{itemize}
    \item [(b)] \textbf{Did you mention the license of the assets?}
        \begin{itemize}
            \item \answerYes{Yes, the licenses of the assets used are mentioned and followed.}
        \end{itemize}
    \item [(c)] \textbf{Did you include any new assets in the supplemental material or as a URL?}
        \begin{itemize}
            \item \answerYes{Yes, upon acceptance, we will release our code and data, and document it in a repository linked via URL.}
        \end{itemize}
    \item [(d)] \textbf{Did you discuss whether and how consent was obtained from people whose data you’re using/curating?}
        \begin{itemize}
            \item \answerNA{NA}
        \end{itemize}
    \item [(e)] \textbf{Did you discuss whether the data you are using/curating contains personally identifiable information or offensive content?}
        \begin{itemize}
            \item \answerYes{Yes, the associated manuscripts address concerns about personally identifiable information.}
        \end{itemize}
    \item [(f)] \textbf{If you are curating or releasing new datasets, did you discuss how you intend to make your datasets FAIR (see FORCE11 (2020))?}
        \begin{itemize}
            \item \answerNA{NA}
        \end{itemize}
    \item [(g)] \textbf{If you are curating or releasing new datasets, did you create a Datasheet for the Dataset (see Gebru et al. (2021))?}
        \begin{itemize}
            \item \answerNA{NA}
        \end{itemize}
\end{itemize}
\end{enumerate}

\section{Ethical Statement} 
Throughout our research process, we have adhered to stringent ethical standards. We have ensured that no personally identifiable information was utilized, and all analyses were conducted on aggregated data. Additionally, we have thoroughly considered the societal impacts of our research. We acknowledge the potential misuse of our findings by malicious entities to target vulnerable individuals susceptible to manipulation. However, we firmly believe that the significance of our work outweighs these risks. Our research has the potential to inform the development of online systems that prioritize safety and responsibility for all individuals, thereby contributing positively to the integrity of digital environments.

\newpage
\section{Appendix}
\subsection{Frequency Distributions of IAR and SAR}
\begin{figure}[ht]
    \centering
    \begin{subfigure}{.5\columnwidth}
        \centering
        \includegraphics[width=\linewidth]{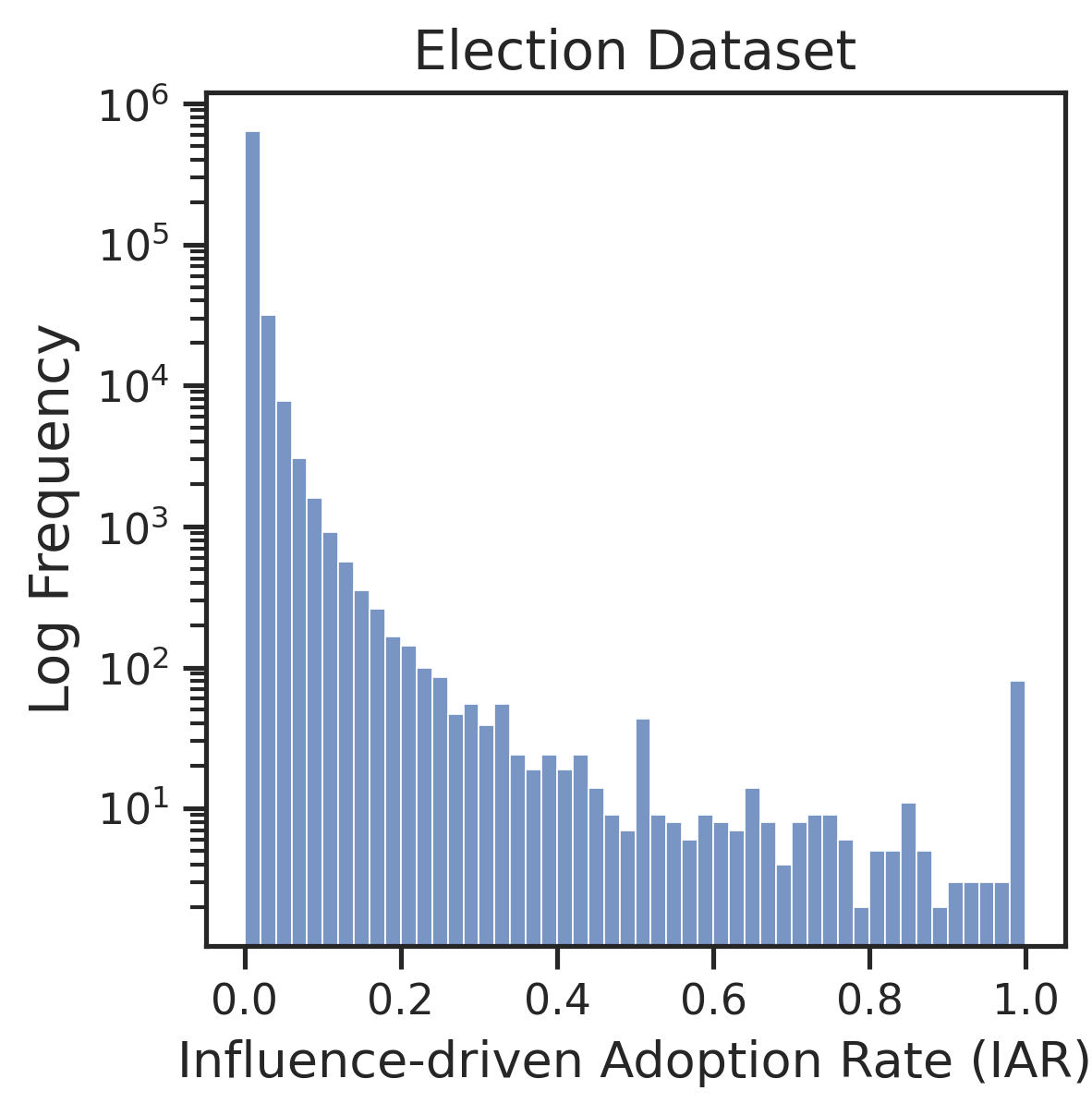}
        \label{IAR_freq_election}
    \end{subfigure}%
    \begin{subfigure}{.5\columnwidth}
        \centering
        \includegraphics[width=\linewidth]{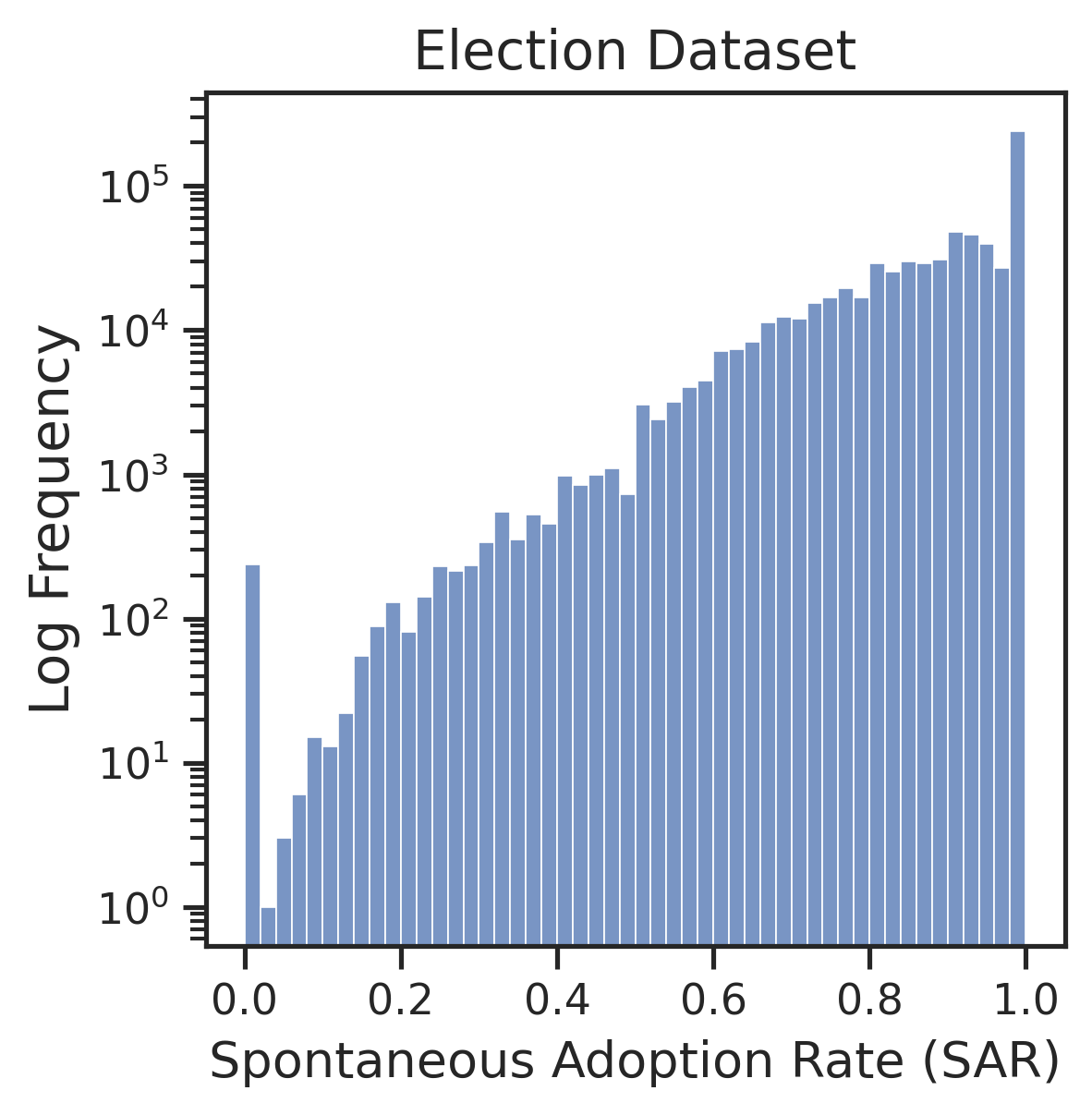}
        \label{SAR_freq_election}
    \end{subfigure}

    \begin{subfigure}{.5\columnwidth}
        \centering
        \includegraphics[width=\linewidth]{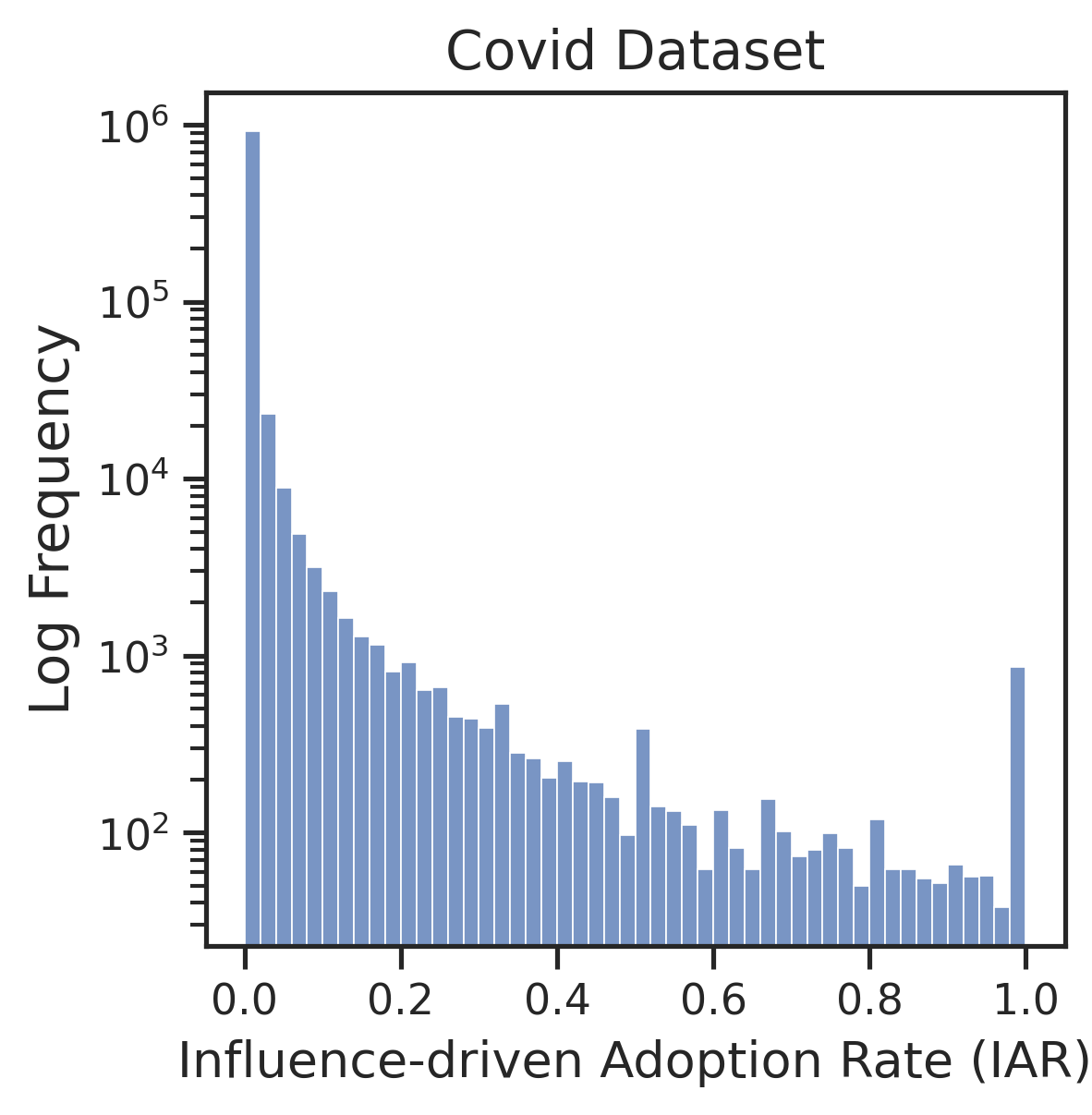}
        \label{IAR_freq_covid}
    \end{subfigure}%
    \begin{subfigure}{.5\columnwidth}
        \centering
        \includegraphics[width=\linewidth]{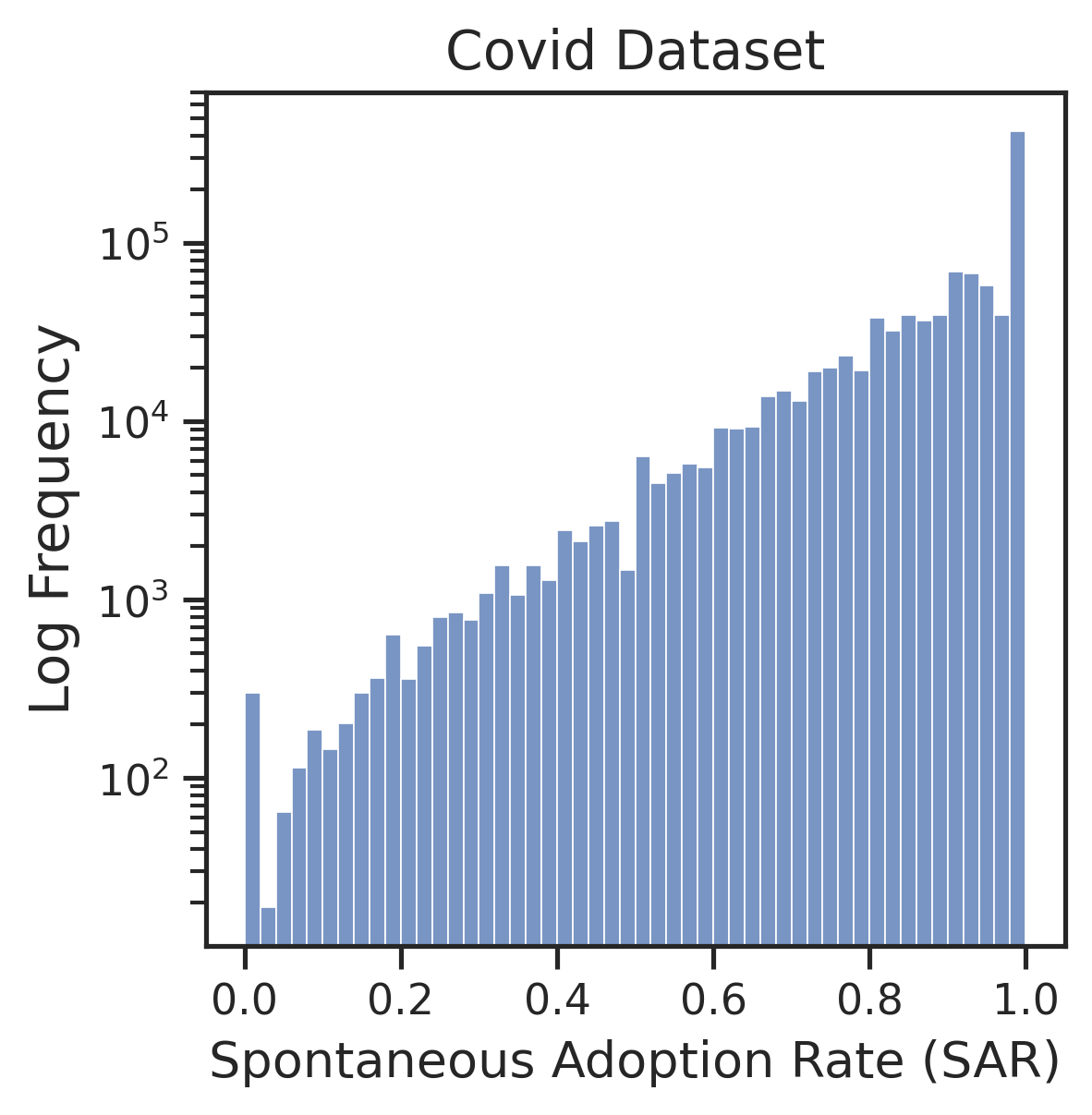}
        \label{SAR_freq_covid}
    \end{subfigure}
    \captionsetup{justification=raggedright,singlelinecheck=false}
    \caption{Frequency distributions of the IAR and SAR metrics in \textsc{Election} and \textsc{Covid} datasets.}
    \label{IAR-SAR-frequency}
\end{figure}

Figure \ref{IAR-SAR-frequency} shows the frequency distributions of the IAR and SAR metrics in \textsc{Election} and \textsc{Covid} datasets.

\subsection{Explanation of the Null Models}
\paragraph{Baseline 1: Edge Swapping}
This null model is designed to test network robustness while preserving the original network's degree distribution and total number of edges. The model operates by implementing an edge swapping mechanism: (1) Pairs of edges are randomly selected within the network. Care is taken to ensure that these selections do not result in multiple edges between the same pair of nodes (unless the original network allows for multiple edges) or self-loops; (2) The endpoints of these selected edge pairs are swapped---this means that if edges $(u, v)$ and $(x, y)$ are chosen, they might be swapped to $(u, y)$ and $(x, v)$; (3) This process is repeated multiple times to ensure thorough randomization while keeping the number of connections (degree) each node has intact.

\paragraph{Baseline 2: Random Neighbor Assignment}
This model tests the robustness of network properties against random reassignment of edges: (1) Each node in the network retains its degree but has its neighbors randomly reassigned. This means that while each node will still have the same number of connections, the actual connections (neighbors) will be randomly chosen; (2) The total number of edges and the overall degrees are preserved, similar to Baseline 1. However, unlike edge swapping, this method completely disrupts any existing neighbor relationships. The objective here is to create a completely randomized structure where only the degree sequence of the original network is preserved.


\subsection{Paradox Holding Probability $p(k, s)$ at Varying Degrees $k$}
\begin{figure*}
    \centering    
    \includegraphics[width=12cm]{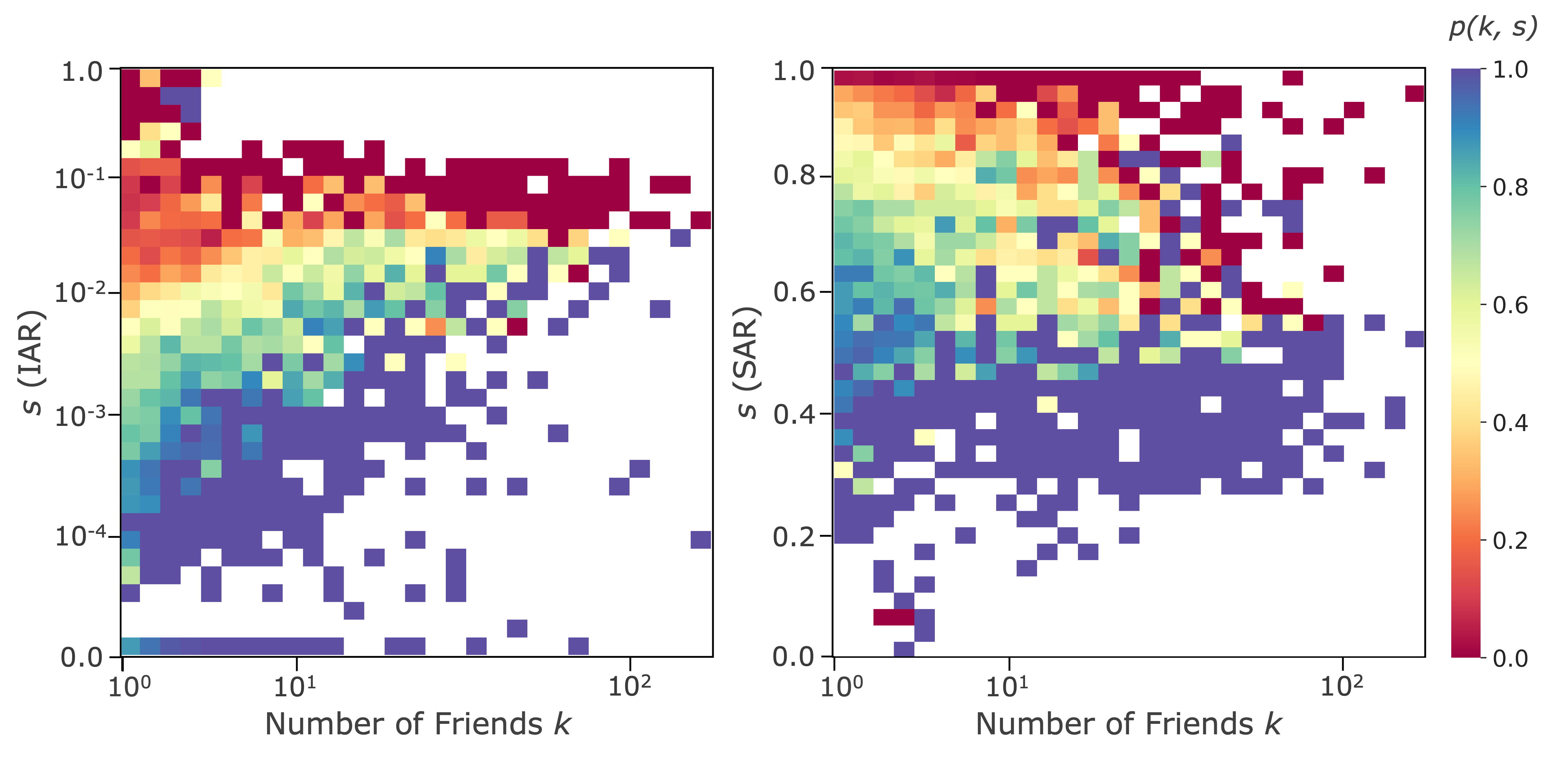} 
    \caption{Paradox holding probability $p(k, s)$ at varying degrees $k$ using the \textit{Interaction} network in the \textsc{Election} dataset.}
    \label{IAR-SAR-heatmap2}
\end{figure*}

Figure \ref{IAR-SAR-heatmap2} depicts the paradox holding probability $p(k, s)$ as a function of degree $k$ and susceptibility metric $s$ using the \textit{Interaction} network in the \textsc{Election} dataset.

\subsection{Random Forest Model Parameters}
Parameters tested included the number of trees (``n\_estimators''; 100 to 1000 in increments of 50), features per split (``max\_features''; scaled linearly from 1 to the total number of features), maximum tree depth (``max\_depth''; 10 to 100 and unlimited), and minimum samples for splits and leaves (``min\_samples\_split'' and ``min\_samples\_leaf''), using values of 2, 5, 10, and 1, 2, 4, respectively. Table \ref{random_forest_parameters} shows a summary of the random forest model parameters.

\begin{table}[H]
  \centering
  \small
  \caption{Summary of the random forest model parameters.}
  \label{random_forest_parameters}
  \begin{tabular}{lcccc}
  \toprule
    & \multicolumn{2}{c}{\textsc{Election}} & \multicolumn{2}{c}{\textsc{Covid}}\\
    \cmidrule(lr){2-3} \cmidrule(lr){4-5}
    \textit{Susceptibility Metric} & IAR & SAR & IAR & SAR\\
    \midrule
    n\_estimators & 750 & 800 & 750 & 800\\
    max\_features & 4 & 4 & 4 & 4\\
    max\_depth & 70 & 80 & 70 & 80\\
    min\_samples\_split & 5 & 10 & 5 & 10\\
    min\_samples\_leaf & 2 & 2 & 2 & 2\\
    \bottomrule
\end{tabular}
\end{table}

\subsection{SHAP Values of IAR and SAR Prediction}

\begin{figure}[H]
    \centering
    \includegraphics[width=0.48\textwidth]{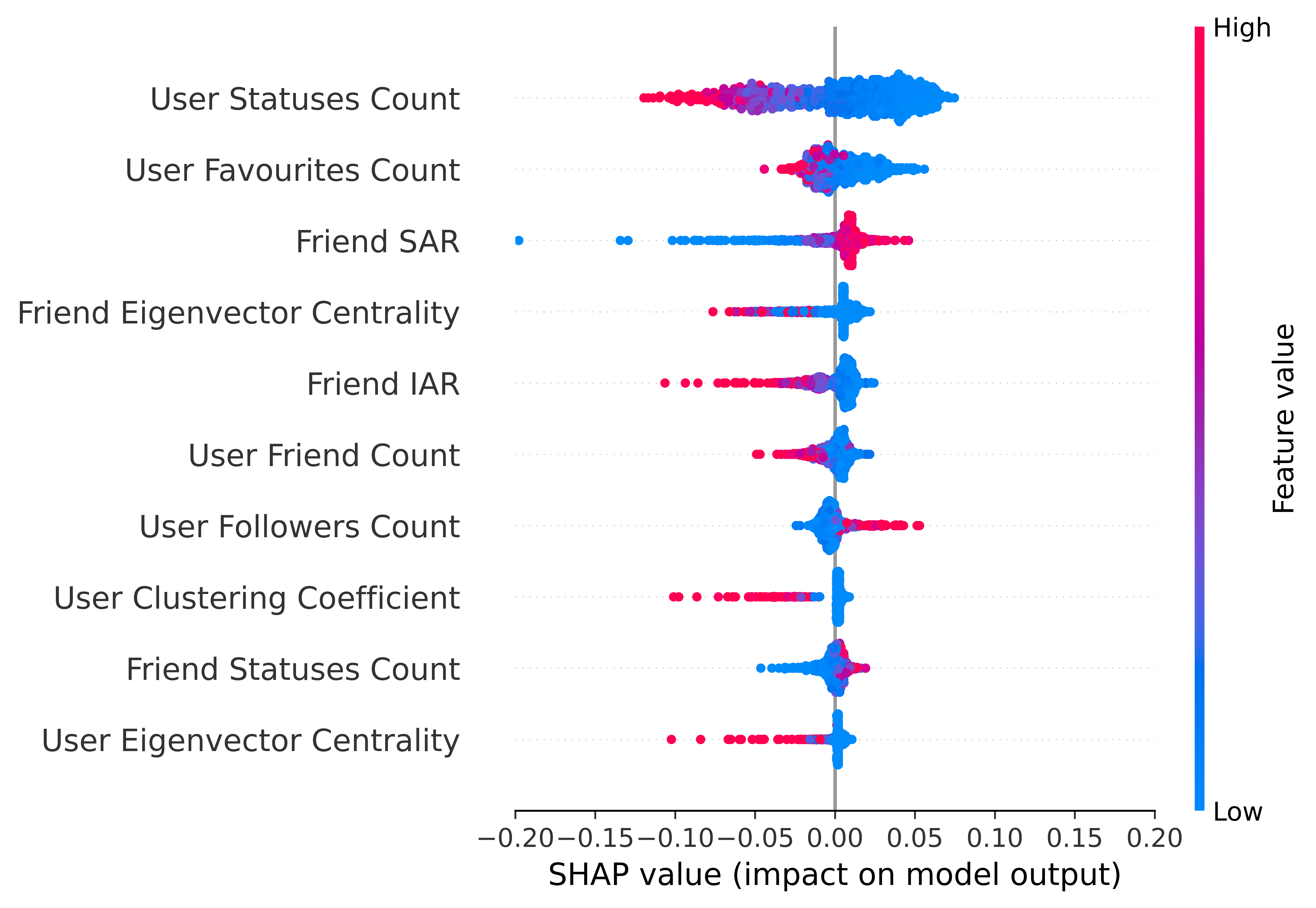}
    \caption{SHAP value distributions for SAR predictions in the \textsc{Election} dataset. Each distribution illustrates the direction and magnitude of the impact of individual features on the model's output, with features ordered by importance from top to bottom.}
    \label{SHAP-election-SAR}
\end{figure}

\begin{figure}[H]
    \centering
    \includegraphics[width=0.48\textwidth]{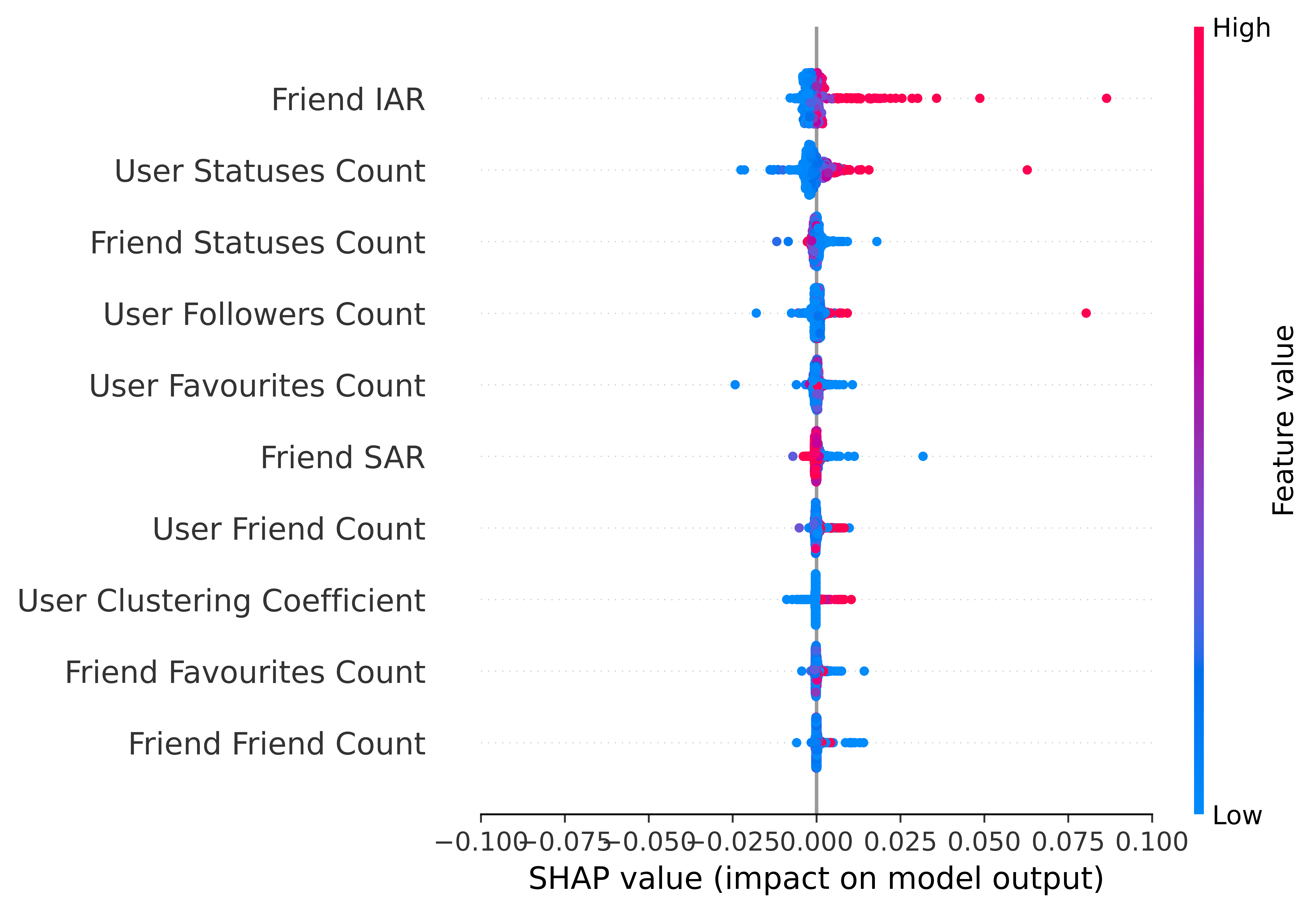}
    \caption{SHAP value distributions for IAR predictions in the \textsc{Election} dataset. Each distribution illustrates the direction and magnitude of the impact of individual features on the model's output, with features ordered by importance from top to bottom.}
    \label{SHAP-election-IAR}
\end{figure}

Figure \ref{SHAP-election-SAR} shows the SHAP value distributions for SAR predictions in the \textsc{Election} dataset. Figure \ref{SHAP-election-IAR} shows the SHAP value distributions for IAR predictions in the \textsc{Election} dataset. 

\subsection{Sensitivity Analysis with Varying Thresholds for Target User Identification}

To address potential sampling biases introduced by the network construction method, we experiment with different URL adoption thresholds to identify target users, specifically testing thresholds of 5 and 20 URLs, in addition to the previously used threshold of 10 URLs. Similarly, we construct three distinct friendship networks based on the types of reciprocal interactions between these target users: \textit{Interaction}, \textit{Retweet}, and \textit{Mention} networks. The resulting node and edge counts for the friendship networks in both datasets, based on these new thresholds, are presented in Table \ref{new_exposure_network}. \footnote{The results for the 10 URL threshold ($G_{10}(V,E)$)is available in the main paper.}

Consistent with previous findings, the results indicate that both susceptibility metrics (IAR and SAR) exhibit homophily within friendship networks, constructed using thresholds of $5$ and $20$ shared URLs (Table \ref{threshold_homophily}). Additionally, IAR demonstrates a stronger degree of homophily compared to SAR. Furthermore, as shown in Table \ref{GFP_threshold}, the \emph{Generalized Friendship Paradox} holds for influence-driven adoption (IAR) but not for spontaneous adoption (SAR) under both thresholds, corroborating our earlier conclusions. These results suggest that our main findings are robust across different sample sizes.

\begin{table}
  \centering
  \small
  \caption{Number of nodes ($|V|$) and edges ($|E|$) in the user networks $G_{5}(V,E)$, $G_{10}(V,E)$, and $G_{20}(V,E)$, constructed with URL thresholds of $5$ and $20$, respectively.}
  \label{new_exposure_network}
  
    \begin{tabular}{lcccc}
  \toprule
    & \multicolumn{2}{c}{\textsc{Election}} & \multicolumn{2}{c}{\textsc{Covid}}\\
    \cmidrule(lr){2-3} \cmidrule(lr){4-5}
    $G_{5}(V,E)$ & $|V|$ & $|E|$ & $|V|$ & $|E|$\\
    \midrule
    Interaction & 98,592 & 242,856 & 134,409 & 300,432\\
    Retweet & 56,701 & 106,893 & 119,456 & 198,654\\
    Mention & 94,702 & 148,114 & 115,584 & 173,022\\
    \bottomrule
    \end{tabular}

    \begin{tabular}{lcccc}
  \toprule
    & \multicolumn{2}{c}{\textsc{Election}} & \multicolumn{2}{c}{\textsc{Covid}}\\
    \cmidrule(lr){2-3} \cmidrule(lr){4-5}
    $G_{10}(V,E)$ & $|V|$ & $|E|$ & $|V|$ & $|E|$\\
    \midrule
    Interaction & 49,794 & 117,384 & 73,960 & 163,974\\
    Retweet & 25,077 & 49,679 & 53,161 & 90,800\\
    Mention & 45,289 & 72,554 & 56,060 & 81,969\\
    \bottomrule
    \end{tabular}
    
  \begin{tabular}{lcccc}
  \toprule
    & \multicolumn{2}{c}{\textsc{Election}} & \multicolumn{2}{c}{\textsc{Covid}}\\
    \cmidrule(lr){2-3} \cmidrule(lr){4-5}
    $G_{20}(V,E)$ & $|V|$ & $|E|$ & $|V|$ & $|E|$\\
    \midrule
    Interaction & 45,845 & 111,402 & 67,702 & 154,068\\
    Retweet & 23,824 & 48,588 & 50,192 & 87,513\\
    Mention & 41,536 & 67,632 & 50,695 & 75,227\\
    \bottomrule
\end{tabular}
\end{table}

\begin{table}
  \centering
  \small
  \caption{The susceptibility metrics (IAR and SAR) exhibit homophily within the social network, as evidenced by the weighted average correlation (Wgt. Avg. Corr.) between users and their friends. All correlation coefficients are significant ($p < 0.001$).}
  \label{threshold_homophily}
  \begin{tabular}{lcccc}
  \toprule
    $G_{5}(V,E)$ & \multicolumn{2}{c}{\textsc{Election}} & \multicolumn{2}{c}{\textsc{Covid}}\\
    \cmidrule(lr){2-3} \cmidrule(lr){4-5}
    Wgt. Avg. Corr. & IAR & SAR & IAR & SAR\\
    \midrule
    Interaction & 0.512 & 0.344 & 0.545 & 0.352\\
    Retweet & 0.520 & 0.371 & 0.569 & 0.356\\
    Mention & 0.458 & 0.302 & 0.346 & 0.288\\
    \bottomrule
\end{tabular}

\begin{tabular}{lcccc}
  \toprule
    $G_{10}(V,E)$ & \multicolumn{2}{c}{\textsc{Election}} & \multicolumn{2}{c}{\textsc{Covid}}\\
    \cmidrule(lr){2-3} \cmidrule(lr){4-5}
    Wgt. Avg. Corr. & IAR & SAR & IAR & SAR\\
    \midrule
    Interaction & 0.541 & 0.356 & 0.564 & 0.369\\
    Retweet & 0.541 & 0.391 & 0.577 & 0.364\\
    Mention & 0.483 & 0.315 & 0.372 & 0.299\\
    \bottomrule
\end{tabular}

  \begin{tabular}{lcccc}
  \toprule
    $G_{20}(V,E)$ & \multicolumn{2}{c}{\textsc{Election}} & \multicolumn{2}{c}{\textsc{Covid}}\\
    \cmidrule(lr){2-3} \cmidrule(lr){4-5}
    Wgt. Avg. Corr. & IAR & SAR & IAR & SAR\\
    \midrule
    Interaction & 0.579 & 0.354 & 0.577 & 0.365\\
    Retweet & 0.586 & 0.389 & 0.589 & 0.367\\
    Mention & 0.506 & 0.324 & 0.399 & 0.294\\
    \bottomrule
\end{tabular}
\end{table}

\begin{table*}[t]
\vspace*{0pt}
  \centering
  \small
  \caption{Empirical results for the \emph{Generalized Friendship Paradox} in three friendship networks derived from the \textsc{Election} dataset are presented. For each susceptibility metric $s$ (IAR and SAR), we calculate the Spearman correlation coefficient $\rho_{ks}$ between degree $k$ and $s$ ($p < 0.001$ for all $\rho_{ks}$), average paradox holding probability $P$ and the average susceptibility $\langle s \rangle$ of nodes and their friends $\langle s \rangle_{nn}$.}
  \label{GFP_threshold}
    \begin{tabular}{@{}lcrcccc@{}}
  \toprule
    \multirow{2}{*}{\textbf{$G_{5}(V,E)$}} & \multirow{2}{*}{\textbf{Metric} $s$} & \multirow{2}{*}{\textbf{$\rho_{ks}$}} & Individual-level GFP & 
    \multicolumn{3}{c}{Network-level GFP} \\
    \cmidrule(lr){4-4} \cmidrule(lr){5-7}
     & & & 
        $P$ & 
    \textbf{$\langle s \rangle$} & & 
        $\langle s \rangle_{nn}$\\
    \midrule
    \multirow{2}{*}{Interaction} & IAR & 0.188 & 0.527 & 0.005 & $<$ & 0.006 \\
    & SAR & $-$0.174 & 0.385 & 0.889 & $>$ & 0.876 \\
    \midrule
    \multirow{2}{*}{Retweet} & IAR & 0.302 & 0.493 & 0.012 & $<$ & 0.013 \\
    & SAR & $-$0.296 & 0.419 & 0.826 & $>$ & 0.812 \\
    \midrule
    \multirow{2}{*}{Mention} & IAR & 0.115 & 0.532 & 0.005 & $<$ & 0.006 \\
    & SAR & $-$0.103 & 0.316 & 0.902 & $>$ & 0.889 \\
    \bottomrule
\end{tabular}

\begin{tabular}{@{}lcrcccc@{}}
  \toprule
    \multirow{2}{*}{\textbf{$G_{10}(V,E)$}} & \multirow{2}{*}{\textbf{Metric} $s$} & \multirow{2}{*}{\textbf{$\rho_{ks}$}} & Individual-level GFP & 
    \multicolumn{3}{c}{Network-level GFP} \\
    \cmidrule(lr){4-4} \cmidrule(lr){5-7}
     & & & 
        $P$ & 
    \textbf{$\langle s \rangle$} & & 
        $\langle s \rangle_{nn}$\\
    \midrule
    \multirow{2}{*}{Interaction} & IAR & 0.179 & 0.541 & 0.007 & $<$ & 0.008 \\
    & SAR & $-$0.166 & 0.409 & 0.880 & $>$ & 0.868 \\
    \midrule
    \multirow{2}{*}{Retweet} & IAR & 0.298 & 0.501 & 0.013 & $<$ & 0.014 \\
    & SAR & $-$0.294 & 0.425 & 0.821 & $>$ & 0.809 \\
    \midrule
    \multirow{2}{*}{Mention} & IAR & 0.111 & 0.544 & 0.005 & $<$ & 0.006 \\
    & SAR & $-$0.099 & 0.383 & 0.896 & $>$ & 0.880 \\
    \bottomrule
\end{tabular}

  \begin{tabular}{@{}lcrcccc@{}}
  \toprule
    \multirow{2}{*}{\textbf{$G_{20}(V,E)$}} & \multirow{2}{*}{\textbf{Metric} $s$} & \multirow{2}{*}{\textbf{$\rho_{ks}$}} & Individual-level GFP & 
    \multicolumn{3}{c}{Network-level GFP} \\
    \cmidrule(lr){4-4} \cmidrule(lr){5-7}
     & & & 
        $P$ & 
    \textbf{$\langle s \rangle$} & & 
        $\langle s \rangle_{nn}$\\
    \midrule
    \multirow{2}{*}{Interaction} & IAR & 0.170 & 0.548 & 0.007 & $<$ & 0.009 \\
    & SAR & $-$0.159 & 0.419 & 0.874 & $>$ & 0.863 \\
    \midrule
    \multirow{2}{*}{Retweet} & IAR & 0.295 & 0.504 & 0.013 & $<$ & 0.014 \\
    & SAR & $-$0.294 & 0.428 & 0.817 & $>$ & 0.805 \\
    \midrule
    \multirow{2}{*}{Mention} & IAR & 0.099 & 0.552 & 0.006 & $<$ & 0.007 \\
    & SAR & $-$0.089 & 0.393 & 0.890 & $>$ & 0.874 \\
    \bottomrule
\end{tabular}
\end{table*}

\end{document}